\documentclass[fleqn]{article}
\usepackage{emlines2,bezier,amscd}
\usepackage{amssymb}
\newcommand{\Inter}[2]{\ensuremath{\mathcal{I}(#1,#2)}}
\newcommand{\Mat}[3]{\ensuremath{\mathrm{M}_{#1\times #2}(#3)}}
\newcommand{\df}{\ensuremath{\mathrm{DEF}_{\mathcal{M}}(\mathcal{B})}}
\newcommand{\vardf}[2]{\ensuremath{\mathrm{DEF}_{#1}(#2)}}
\newcommand{\prf}{{\bf Proof.\ }}
\newcommand{\de}[2]{\ensuremath{(\iota_{#1},\iota_{#2})}}
\newcommand{\depr}[2]{\ensuremath{(\iota'_{#1},\iota'_{#2})}}
\newcommand{\tm}[2]{\ensuremath{#1\times #2}}
\newcommand{\zt}[2]{\ensuremath{\iota_{#2}\circ\iota_{#1}^{-1}}}
\newcommand{\Diff}[1]{\ensuremath{{\rm Diff}(#1)}}

\newcommand{\Prj}[1]{\ensuremath{{\rm Pr}_{#1}}}

\newcommand{\proj}[2]{\ensuremath{\pi_{#1}(#2)}}
\newcommand{\hmt}[2]{\ensuremath{#1\stackrel{\mathrm{H}}{\sim} #2}}

\newcommand{\mt}[1]{\ensuremath{\mathrm{MOT}_{\mathcal{M}}(#1)}}
\newcommand{\varmt}[2]{\ensuremath{\mathrm{MOT}_{#1}(#2)}}
\newcommand{\emb}[2]{\ensuremath{{\rm Emb}(#1,#2)}}
\newcommand{\Img}[1]{\ensuremath{{\rm Im}(#1)}}
\newcommand{\Dom}[1]{\ensuremath{{\rm Dom}(#1)}}
\newtheorem{pred}{Proposition}
\newcommand{\Hs}[1]{\ensuremath{{\rm Hist}(#1)}}
\newcommand{\hs}[2]{\ensuremath{F_{(#1)#2}}}
\newcommand{\dst}[9]
{\ensuremath{\mathfrak{#1}#2=\langle\mathcal{#3}#4,\mathcal{#5}#6,\mathcal{#7}#8,#9\rangle}}

\title{DEFORMATIONAL STRUCTURES ON SMOOTH MANIFOLDS}
\author{Sergey S.Kokarev\thanks{sergey@yspu.yar.ru}}
\date{Department of theoretical physics, r.409, YSPU,
                       Respublikanskaya 108, Yaroslavl, 150000, Russia}

\begin{document}
\maketitle

\begin{abstract}
\small Abstract deformational structures, in many aspects generalizing standard
elasticity theory, are investigated. Within free deformational structures we
define algebra of deformations, classify them by its special properties, define
motions and conformal motions together with  deformational decomposition of
manifolds, generalizing isometry of Riemannian spaces and consider  some
physical examples. In frame of dynamical deformational structures we formulate
variational procedure for evolutional and static cases together with boundary
conditions, derive dynamical (equilibrium in static case) equations, consider
perturbative approach and perform deformational realization of the well known
classical field-theoretical  topics: strings and branes theories, classical
mechanics of solids, gravity and  Maxwell electrodynamics.
\end{abstract}

\vspace{0.2cm} {\small PACS: 04.50.+h; 45.40.-f; 46.25.-y;
02.40.Hw}\vspace{0.5cm} ]

\section{Introduction}\label{intro}

Recent time the strong tendency to inclusion of  embedded objects into the
scope of  theoretical and mathematical physics is observed (see references in
\cite{pav}). We should relate to the subject all strings and branes models
\cite{green,plectures}, including their supersymmetric and noncommutative
generalizations \cite{noncomm}, embedding methods of GR  \cite{kobayasi} and
its alternative formulations and generalizations \cite{pav1}, geometrical
methods of nonlinear differential equations theory and jets approach
\cite{vinogr} and many other things. Probably, such central position of the
"embedded objects" in modern physics can't be accidental:  it may reflect
either multidimensional nature of physical reality, observed through all its
levels, or some "immanent" for us, as observers, means for its description.

At the same time,  majority of the  field-the\-o\-re\-ti\-cal models,
exploiting embedded objects, reveal amazing and, in our opinion, deep
interrelations with some general ideas of elasticity theory of continuous media
\cite{land} may be with a number of "nonstandard" properties such as
nonlinearity, plasticity, viscosity, anisotropy, internal spin, nematic or
smectic structures or memory \cite{born,tar,hehl,dmitriev,visser,un}.
Particularly, in papers \cite{kok1,kok2,kok3,kok4,kok5} it has been shown, that
Einstein GR and standard classical solids dynamics admit natural formulation in
terms of mechanical straining of thin 4D plates and 4D strings (strongly tensed
bars) respectively.

Interesting and important problem, arising under such unifying of embedding and
elasticity ideas, is to extract and formulate general ideas of continuous media
physics in its the most abstract and general form, independent on peculiarities
of one or another theory. So, we intend to follow  the line of investigations,
which can be called {\it general theory of deformational structures
($d$-structures)} with the aim
--- to formulate and work out universal language for the objects, which are
able, in some sense, to be "deformed".

Although we'll restrict ourself by the case of real manifolds, majority of
statements will take place after suitable complex generalization, which is
necessary for constructing of  quantum $d$-structures. Moreover, some general
concepts "survive" even without smooth structures, but we reserve the more
abstract schemes for future.

Present paper\footnote{This is revised and essentially more developed version
of \cite{kok6}, which is, in turn, small part of talk, presented at 5-th
Asian-Pacific conference (Moscow, October 2001).} is devoted to some first
principles of this program. We work out "deformational terminology" and set
some general propositions, statements and relations, which can be recognized
within well known theories and which can be used in future works.

Within the first half of the paper (Sec.\ref{freestr}) we consider free
$d$-structures, generalizing kinematics of standard elasticity theory and
reflecting, mainly, geometrical properties of  a number of physical models. The
second half  (Secc.\ref{dyn},\ref{eqsc},\ref{exam}) is devoted
 to dynamical $d$-structures, which
include, apart from kinematics, some dynamical principle and reflect, mainly,
physical properties of field-theoretical  models. Examples of deformational
structures, performed in the paper, involve elasticity theory together with its
(generally-)covariant generalization, Hamiltonian formalism, bundle spaces with
invariant connection, thermodynamics, strings and branes theories, classical
solids dynamics, gravity, Maxwell electrodynamics.
 Some more subtle technical questions are investigated  in Appendixes.

Always, when it is possible we use standard notations of sets theory
\cite{scorn}, smooth manifolds theory \cite{warner} and
(almost\footnote{Coordinates in ambient space we denote by small Latin letters
with big Latin indexes
--- $x^A$ $A=1,\dots n,$ on embedded $d$-object ---  by Greek letters with Greek
indexes --- $\xi^\alpha,$ $\alpha=1,\dots,d$. Such doubling is useful for
coordinateless symbolic notations.} anywhere) use  coordinateless
representation of tensor equations. Particularly, we'll denote by

${\rm Dom},\ {\rm Im}$ --- domains and images of mappings;

$\stackrel{\rho}{\sim}$ --- equivalence relation $\rho$;

${\rm Dg}(A\times A)$ --- diagonal of a direct product (i.e. set of pairs
$(a,a)\in A\times A$);

$\pi_\rho$ --- mapping on quotient space with respect to equivalence $\rho$;

$[a]_\rho$ --- class of equivalence of the element $a$ with respect to $\rho$;

$A\le B$ --- $A$ is sub(pseudo)group of (pse\-u\-do)\-gro\-up $B$;

$\partial_x X=X_{|x}\equiv\partial X/\partial x$ --- partial derivative;

$T(r,s)$ --- space of tensors of covariant valency $r$ and contravariant
valency $s$;

$(\ ,\ )$ --- scalar product in different tensor spaces;

$\langle\ ,\ \rangle$ --- pairing of tensors and linear functionals over them;

$\mathrm{M}_{m\times n}(\mathfrak{C})$ module of $m\times n$ matrices over ring
$\mathfrak{C};$

${\rm Hom}(A,B)$ --- space of linear mappings of modules (linear spaces) $A\to
B.$

\section{Free deformational structures}\label{freestr}

\subsection{Definitions}\label{defree}

We call {\it free deformational structure  $\mathfrak{D}$} the collection
$\langle\mathcal{B},\mathcal{M},\mathcal{E},\Theta\rangle,$ where:

$\mathcal{B}$ and $\mathcal{M}$ --- smooth, connected, closed manifolds, ${\rm
dim}\,\mathcal{B}=d,$ ${\rm dim}\,\mathcal{M}=n\ge d;$

$\mathcal{E}\subseteq\emb{\mathcal{B}}{\mathcal{M}}$ --- some subset of all
smooth embeddings $\mathcal{B}\hookrightarrow\mathcal{M};$

$\Theta\in\Omega^{\otimes p}(\mathcal{M})$ --- some smooth real-valued form of
degree $p$ on $\mathcal{M}.$ In what follows we'll call: $\mathcal{B}$
--- {\it $d$-body},  $\mathcal{M}$ --- {\it $d$-manifold,}
$\Theta$
--- {\it $d$-metrics,} and image
%\[
%$\begin{equation}\label{emb}
$\iota(\mathcal{B})\equiv\mathcal{S}\subseteq\mathcal{M}$
%\end{equation}
%\]
for some $\iota\in\mathcal{E}$ --- {\it $d$-object} or {\it deformant.}

Any embedding $\iota$ induces form $(d\iota)^\ast\Theta\in\Omega^{\otimes
p}(\mathcal{B}),$ where $(d\iota)^\ast$
--- embedding $\iota$ codifferential\footnote{We denote by $(d\iota)^\ast$ mappings
$\Omega^{\otimes p}(\mathcal{M})\to\Omega^{\otimes p}(\mathcal{B})$ for any
$p.$}, mapping $\Omega^{\otimes p}(\mathcal{M})\to\Omega^{\otimes
p}(\mathcal{B}).$ Let consider some another embedding $\iota'\in\mathcal{E},$
which induces its own $d$-object
$\iota'(\mathcal{B})\equiv\mathcal{S'}\subseteq\mathcal{M}.$ In
$\Omega^{\otimes p}(\mathcal{B})$ we'll have the form $(d\iota')^\ast\Theta.$
Easily to see, that the composition
\begin{equation}\label{def}
\iota'\circ\iota^{-1}\equiv\zeta
\end{equation} is diffeomorphism
$\mathcal{S}\to\mathcal{S'}=\zeta(\mathcal{S}),$  which we'll call {\it
deformation} of $d$-body in $\mathcal{M}$.

Any deformation $\zeta$ has natural local measure
---  difference of two forms, taken at the same point $b\in\mathcal{B}$:
\[
(d\iota')^\ast\Theta(b)-(d\iota)^\ast\Theta(b)\equiv \Delta_{\mathcal{B}}(b),
\] where we have introduced notation $\Delta_{\mathcal{B}}$ for {\it deformation
form} on $\mathcal{B}.$ Using definition (\ref{def}) and the well known
composition property of codifferential:
\begin{equation}\label{codiff}
(d(\alpha\circ\beta))^\ast=(d\beta)^\ast\circ (d\alpha)^\ast,
\end{equation}
 we  obtain the
equivalent representation:
\begin{equation}\label{formb1}
\Delta_{\mathcal{B}}=(d\iota)^\ast ((d\zeta)^\ast \Theta-\Theta),
\end{equation}
and define the deformation form
\begin{equation}\label{forms}
\Delta_{\mathcal{S}}\equiv
((d\iota)^\ast)^{-1}\Delta_{\mathcal{B}}=(d\zeta)^\ast\Theta-\Theta
\end{equation}
on the deformant $\mathcal{S}.$ Note, that the representations (\ref{formb1})
and (\ref{forms}) correspond to material and referent descriptions of
deformable bodies in classical continuum media dynamics \cite{trus}.

\subsection{Algebra of deformations}\label{pseudo}

For any deformation $\zeta$ let define the subsets of $\mathcal{E}:$
\[
\Prj{1}(\zeta)=\{\iota\in\mathcal{E}\ |\ \Img{\iota}=\Dom{\zeta}\};\ \
\Prj{2}(\zeta)=\{\iota\in\mathcal{E}\ |\ \Img{\iota}=\Img{\zeta}\}.
\]
As it follows from the definition (\ref{def}), the set of all deformations of
the $d$-body in $\mathcal{M}$, which we'll denote
$\mathrm{DEF}_{\mathcal{M}}(\mathcal{B}),$ can be treated as image of the
surjective map $\phi:\ \mathcal{E}\times\mathcal{E}\to\df,$ acting by the rule:
\begin{equation}\label{rule}
\phi(\iota_\alpha,\iota_\beta)=\iota_\beta\circ\iota_\alpha^{-1}\equiv\zeta_{\alpha\beta}.
\end{equation}
The following proposition clears the relation between
$\tm{\mathcal{E}}{\mathcal{E}}$ and $\df.$
\begin{pred}\label{corresp}\ Fibre
$\phi^{-1}(\zeta)=\{d\in\mathcal{E}\times\mathcal{E}\,|\ d=(\iota_\zeta\circ
l,\zeta\circ\iota_\zeta\circ l)\},$ where $\zeta$
--- some element of $\df,$ $l$ runs all elements from the
${\rm Diff}(\mathcal{B}),$
 and embedding $\iota_\zeta\in\Prj{1}(\zeta).$
\end{pred}

\prf The inclusion $(\iota_\zeta\circ l,\zeta\circ\iota_\zeta\circ
l)\in\phi^{-1}(\zeta)$ immediately follows from the (\ref{rule}). Let the two
elements $\de{\alpha}{\beta}$ and $\de{\gamma}{\delta}$ of
$\tm{\mathcal{E}}{\mathcal{E}}$ defines the same deformation
$\zeta=\zt{\alpha}{\beta}=\zt{\gamma}{\delta}.$ Images of the firsts --
$\iota_\alpha,\ \iota_\gamma$ and of the seconds -- $\iota_\beta,\
\iota_\delta$ embeddings pair-wisely coincide in $\mathcal{M}$ (as domains and
images of the same deformation  $\zeta$ in $\mathcal{M}$ respectively), i.e.:
\[ \iota_{\alpha}(\mathcal{B})=\iota_\gamma(\mathcal{B})=\mathcal{S}\ \
\mbox{\rm and}\ \
\iota_{\beta}(\mathcal{B})=\iota_\delta(\mathcal{B})=\mathcal{S}'. \] Then,
particularly, it follows, that $\iota_\gamma=\iota_\alpha\circ l,$ where $l$
--- some diffeomorphism of the  $d$-body $\mathcal{B}.$ So, if the pairs
$\de{\alpha}{\beta}$ and $\de{\gamma}{\delta}$ lie in the same fiber
$\phi^{-1}(\zeta),$ then they necessary have the form $(\iota_\alpha,
\zeta\circ\iota_\alpha)$ and $(\iota_\alpha\circ l,\zeta\circ\iota_\alpha\circ
l)$ respectively. Simultaneousity  of the two inclusions proves the
proposition.$\square$

% Note, that the set $\Diff{\mathcal{B}}$ can be viewed as subgroup of ${\rm
%Aut}(\mathcal{E}),$ since $\mathcal{E}\circ\Diff{\mathcal{B}}=\mathcal{E}.$
The map $\phi$ endows $\tm{\mathcal{E}}{\mathcal{E}}$ the canonical equivalence
$\stackrel{D}{\sim}$:
$\de{\alpha}{\beta}\stackrel{D}{\sim}\de{\gamma}{\delta},$ if
$\phi\de{\alpha}{\beta}=\phi\de{\gamma}{\delta}$ and we can identify $\df$ with
quotient space\footnote{Such quotient space is sometimes called {\it twisted
multiplication} and in our case is denoted
$\mathcal{E}\times_{\Diff{\mathcal{B}}}\mathcal{E}.$}
$\pi_D(\tm{\mathcal{E}}{\mathcal{E}}),$ consisting of classes
$[\de{\alpha}{\beta}]_D=\zeta_{\alpha\beta},$ such that
$\pi^{-1}_D(\zeta_{\alpha\beta})$ is  the  fiber of Proposition 1, containing
the element $\de{\alpha}{\beta}\in\tm{\mathcal{E}}{\mathcal{E}}.$

 On the set $\pi_D({\tm{\mathcal{E}}{\mathcal{E}}})$ one can introduce
the following binary relation: \[
\rho=\{(\zeta_1,\zeta_2)\in\tm{\pi_D({\tm{\mathcal{E}}{\mathcal{E}}})}{\pi_D({\tm{\mathcal{E}}{\mathcal{E}}})}\,|\,
\Prj{2}(\zeta_1)=\Prj{1}(\zeta_2)\}.
\]
It is easily checked, that $\rho$ is $\rm T-$reflective  and $\rm
T-$an\-ti\-sym\-met\-ric, i.e. $(\zeta,\zeta^{\rm T})\in\rho,$ and, if
simultaneously $(\zeta_{1},\zeta_{2})\in\rho$  and
$(\zeta_{2},\zeta_{1})\in\rho,$  then $zeta_2=\zeta_1^{\rm T}.$ Here
$(\zeta^{\rm T})_{\alpha\beta}\equiv \zeta_{\beta\alpha}.$ We'll call this
relation {\it $\rm T-$to\-ur\-na\-ment\footnote{Tournament is reflective and
antisymmetric binary relation \cite{scorn}.}.}

%Let $\zeta\in \pi_D({\tm{\mathcal{E}}{\mathcal{E}}}).$
Lets denote  $Y^{\mp}_\zeta$ the following subsets:
\[Y^{-}_\zeta\equiv\{\zeta'\in\pi_D({\tm{\mathcal{E}}{\mathcal{E}}})\,|\
(\zeta',\zeta)\in\rho\};\ \
Y^{+}_\zeta\equiv\{\zeta'\in\pi_D({\tm{\mathcal{E}}{\mathcal{E}}})\,|\
(\zeta,\zeta')\in\rho\}.
\]

\begin{pred}\label{ps}
On the set  $\pi_D({\tm{\mathcal{E}}{\mathcal{E}}})$ with  $\rm
T-$tour\-na\-ment $\rho$ there exists pseudogroup structure \footnote{Let
remind, that pseudogroup is a set of elements $\mathcal{A}$, for which
composition $\ast$ is defined may be on some subset (binary relation)
$\mathcal{U}\subset\tm{\mathcal{A}}{\mathcal{A}}$ and where  the following
properties are hold: associativity,  for every $a\in\mathcal{A}$ there exist
unique left $e_a^{-}$ and right $e_a^{+}$ units elements (generally speaking
depending on $a$), lying in $\mathcal{A}$ and there exists unique inverse
element $a^{-1},$ lying in $\mathcal{A},$ such that $e^{-}_a\ast a =a\ast
e^{+}_a =a$ and $a\ast a^{-1}=e^{-}_a,\ \ a^{-1}\ast a=e^{+}_a$ \cite{scorn}.}.
\end{pred}
\prf\ For any $\zeta\in\pi_D({\tm{\mathcal{E}}{\mathcal{E}}})$ and for all
$\zeta'\in Y^{-}_\zeta$ ¨ $\zeta''\in Y^{+}_\zeta$ we define left and right
pseudogroup multiplications as compositions of deformations:
\[\zeta'\ast \zeta\equiv \zeta'\circ \zeta,\ \ \mbox{\rm and }\ \  \zeta\ast \zeta''\equiv
\zeta\circ \zeta''\] respectively. In components:
\[\zeta'\ast
\zeta =[(\iota',\iota_1)]_D\ast[\de{1}{2}]_D\equiv[(\iota',\iota_2)]_D;\ \
\zeta\ast
\zeta''=[\de{1}{2}]_D\ast[(\iota_2,\iota'')]_D\equiv[(\iota_1,\iota'')]_D.\]
Units elements will be given by the expressions:
\[e^{-}_\zeta\equiv[(\iota_\zeta,\iota_\zeta)]_D\in
\pi_D({\rm Dg}({\tm{\mathcal{E}}{\mathcal{E}}})),\ \
e^{+}_\zeta\equiv[(\iota'_\zeta,\iota'_\zeta)]_D\in \pi_D({\rm
Dg}({\tm{\mathcal{E}}{\mathcal{E}}})),\] where $\iota_\zeta\in\Prj{1}(\zeta),$
$\iota'_\zeta\in\Prj{2}(\zeta).$ Finally, for every
$\zeta\in\pi_D({\tm{\mathcal{E}}{\mathcal{E}}})$ there exist unique inverse
element $\zeta^{-1}$ and it is easily to check in components, that
$\zeta^{-1}=\zeta^{\rm T}.$

So, {\it the set $\pi_D(\mathcal{E}\times\mathcal{E})\equiv\df$ ---
pse\-u\-do\-gro\-up.}$\square$

\subsection{Classification of deformations and Boolean matrix
calculus}\label{class}

Lets consider the following formal object:
\[
\mathcal{I}\equiv \left(
\begin{array}{ll} {\rm Dom}\cap{\rm Dom}\ &\ {\rm Dom}\cap{\rm Im}\\
{\rm Im}\cap{\rm Dom}\ &\  {\rm Im}\cap{\rm Im}
\end{array} \right).
\]
It can be understood as the mapping: $\tm{\df}{\df}\to
\Mat{2}{2}{\mathfrak{B}(\mathcal{M})},$ where
$\Mat{2}{2}{\mathfrak{B}(\mathcal{M})}$ --- module of $2\times 2$ matrices over
ring of subsets of $\mathcal{M},$ which form boolean algebra
$\mathfrak{B}(\mathcal{M}).$ For every pair
$(\zeta_1,\zeta_2)\in\tm{\df}{\df},$ such that $\zeta_1:\
\mathcal{S}_1\to\mathcal{S}'_1$ and $\zeta_2:\ \mathcal{S}_2\to\mathcal{S}'_2$
we have:
\[
\mathcal{I}(\zeta_1,\zeta_2)=\left(\begin{array}{cc}
\mathcal{S}_1\cap\mathcal{S}_2\ &\ \mathcal{S}_1\cap\mathcal{S}'_2\\
\mathcal{S}'_1\cap\mathcal{S}_2\ &\ \mathcal{S}'_1\cap\mathcal{S}'_2
\end{array}\right).
\]
We'll call $\mathcal{I}(\zeta_1,\zeta_2)$ {\it matrix of intersection of
$\zeta_1$ and $\zeta_2$.} For  $\zeta_1=\zeta_2=\zeta$ we'll call
$\mathcal{I}(\zeta,\zeta)$ {\it matrix of self-intersection of $\zeta$.} Easily
to check, that matrix of intersection is degenerate on any pair of deformations
in boolean sense, i.e. ${\rm
det}\,\mathcal{I}(\zeta_1,\zeta_2)\equiv\varnothing,$ where determinant is
defined as usually, but  calculation are carried out with the help of boolean
operations $\cap,\setminus.$

The first step to classification of deformations is based on the kind of matrix
$\mathcal{I}(\zeta,\zeta).$ We'll say, that  deformation $\zeta:\
\mathcal{S}\to\mathcal{S}'$

--- is {\it parallel,} if $\Inter{\zeta}{\zeta}$ --- diagonal in boolean sense (i.e.
nondiagonal components are $\varnothing$);

--- is {\it sliding,}\footnote{It is useful to differ the following particular cases:
{\it total sliding,} if $\zeta$ --- sliding with $\mathcal{S}=\mathcal{M}$ and
{\it empty sliding}, if $\zeta$ --- sliding with $\mathcal{S}=\varnothing.$}
 if $\Inter{\zeta}{\zeta}=\mathcal{S}\cdot\Omega,$
where
\[
\Omega\equiv\left(\begin{array}{cc} \mathcal{M}\ &\ \mathcal{M}\\ \mathcal{M}\
&\ \mathcal{M}
\end{array}\right)
\]
and multiplication on "number" $\mathcal{S}$ is  component-wise boolean
multiplication $\cap$ of $\mathcal{S}$ on elements of $\Omega;$

--- is {\it stretch of $\mathcal{S}$,} if
\[
\mathcal{I}(\zeta,\zeta)=\left(\begin{array}{cc} \mathcal{S}\ &\ \mathcal{S}\\
\mathcal{S}\ &\ \mathcal{S}'
\end{array}\right);
\]

--- is {\it contraction of $\mathcal{S}$,} if
\[
\mathcal{I}(\zeta,\zeta)=\left(\begin{array}{cc} \mathcal{S}\ &\ \mathcal{S}'\\
\mathcal{S}'\ &\ \mathcal{S}'
\end{array}\right).
\]

We'll denote this the {\it simplest} classes of  deformations as \[{\rm
Simp}\equiv\{{\rm Par}(\mathcal{S}),\ {\rm Sl}(\mathcal{S}),\ {\rm
Str}(\mathcal{S}),\ {\rm Ctr}(\mathcal{S})\}\] respectively, omitting sometimes
argument $\mathcal{S}.$ We observe,   that by the definitions
\[{\rm
Sl}(\mathcal{M})\cong\vardf{\mathcal{M}}{\mathcal{M}}\equiv\Diff{\mathcal{M}},\
\
 {\rm Str}(\mathcal{S})\cap{\rm Ctr}(\mathcal{S})={\rm Sl}(\mathcal{S}).
\]
Easily to see, that for every $\mathcal{S}=\iota(\mathcal{B}),$ ${\rm
Sl}(\mathcal{S}),$ ${\rm Str}(\mathcal{S}),$ ${\rm Ctr}(\mathcal{S})$ form
subpseudogroups\footnote{The set $\mathcal{A}'\subset\mathcal{A}$ is said to be
subpseudogroup of pseudogroup $\mathcal{A},$ if $\mathcal{A}'$ --- pseudogroup
with respect to composition in $\mathcal{A}.$ We leave notation
$\mathcal{A}'\le\mathcal{A}$ from groups theory \cite{scorn}.} of $\df,$ while
${\rm Par}(\mathcal{S})$ generally speaking, doesn't. Obviously, boolean matrix
calculus become trivial for $\Diff{\mathcal{M}}$, since it is mapped into a
single self-intersection (in fact, intersection too) matrix $\Omega.$

Let $\df\ni\zeta:\ \mathcal{S}\to\mathcal{S}'$ and let
$\mathcal{S}\cap\mathcal{S}'\equiv\mathcal{S}_0$ is connected. Then we can
define deformations $\zeta_{\pm}$ by the rules:
\[
\zeta_{+}:\ \mathcal{S}_0\to\zeta(\mathcal{S}_0)\equiv\mathcal{S}'_0; \ \
\zeta_{-}:\ \mathcal{S}_0\to\zeta^{-1}(\mathcal{S}_0)\equiv\mathcal{S}''_0.
\]
We'll call $\mathcal{S}_0$ --- {\it zeroth self-intersection}, $\zeta_{+}$ ---
{\it first direct} and $\zeta_{-}$ --- {\it first reverse  continuations of
$\zeta$}. Then we introduce the {\it first direct}
$\mathcal{S}_0\cap\mathcal{S}'_0=\mathcal{S}_+$ and {\it first reverse
$\mathcal{S}_0\cap\mathcal{S}''_0=\mathcal{S}_-$ intersections} and {\it second
direct} and {\it reverse continuations of $\zeta$ --- deformations}
$\zeta_{\pm\pm}:$
\[ \zeta_{++}:\ \mathcal{S}_+\to\zeta(\mathcal{S}_+);\ \ \zeta_{+-}:\
\mathcal{S}_+\to\zeta^{-1}(\mathcal{S}_+);\ \ \zeta_{-+}:\
\mathcal{S}_-\to\zeta(\mathcal{S}_-);\ \ \zeta_{--}:\
\mathcal{S}_-\to\zeta^{-1}(\mathcal{S}_-).
\]
Assuming connectedness of $\mathcal{S}_{\pm}$ and continuing this procedure, we
obtain the chain of self-intersections and corresponding chain of deformational
continuations:
\[\mathcal{S}_0\stackrel{\{\zeta_\pm\}}{\to}\{\mathcal{S}_\pm\}\stackrel{\{\zeta_{\pm\pm}\}}{\to}
\dots\stackrel{\{\zeta_{\alpha_{n-1}}\}}{\to}\{\mathcal{S}_{\alpha_{n-1}}\}\stackrel{\{\zeta_{\alpha_{n}}\}}{\to}\{\mathcal{S}_{\alpha_n}\}\dots,\]
where $\{\alpha_n\}$ denotes collection of $2^n$ binary codes of length $n$ of
the kind $i_1i_2\dots i_n,$ $i_k=+,-$. For example, if $\mathcal{S}_{\alpha_n}$
--- some connected fixed $n$-th self-intersection, then we define by induction:
\[\{\zeta_{\alpha_{n+1}}\}\ni\zeta_{\alpha_n+}:\
\mathcal{S}_{\alpha_n}\to\zeta(\mathcal{S}_{\alpha_n});\ \
\{\zeta_{\alpha_{n+1}}\}\ni\zeta_{\alpha_n-}:\
\mathcal{S}_{\alpha_n}\to\zeta^{-1}(\mathcal{S}_{\alpha_n});\]
\[\mathcal{S}_{\alpha_n+}=\mathcal{S}_{\alpha_n}\cap\zeta(\mathcal{S}_{\alpha_n});\
\
\mathcal{S}_{\alpha_n-}=\mathcal{S}_{\alpha_n}\cap\zeta^{-1}(\mathcal{S}_{\alpha_n}).\]
Also we get the set of matrices of $n$-th self-in\-ter\-sec\-ti\-ons as
$\{\mathcal{I}_{\alpha_n}\}\equiv\{\mathcal{I}(\zeta_{\alpha_n},\zeta_{\alpha_n})\}.$

The following two propositions are basic for classifying of intersected
$d$-objects.

\begin{pred}\label{stat}
If $\zeta_{\alpha_n}\in{\rm Simp},$ then all continuations of
$\zeta_{\alpha_n}$ lie in ${\rm Simp}.$
\end{pred}

\prf Let $\zeta_{\alpha_n}\in{\rm Par},$ then
$\mathcal{S}_{\alpha_n}=\varnothing$ and all
$\mathcal{I}_{\alpha_m}=\varnothing_{2\times2}$ for $m>n,$ so
$\zeta_{\alpha_m}\in{\rm Sl}(\varnothing)\equiv{\rm Par}(\varnothing).$

Let $\zeta_{\alpha_n}\in {\rm Sl}(\mathcal{S}_{\alpha_{n-1}}),$  then
$\mathcal{S}_{\alpha_n}=\mathcal{S}_{\alpha_{n-1}}.$ So, we have
$\zeta_{\alpha_n\pm}:\ \mathcal{S}_{\alpha_{n-1}}\to\mathcal{S}_{\alpha_{n-1}}$
and
$\mathcal{I}_{\alpha_m}(\zeta,\zeta)=\mathcal{I}_{\alpha_n}(\zeta,\zeta)=\mathcal{S}_{\alpha_{n-1}}\cdot\Omega$
for all $m\ge n.$

Let $\zeta_{\alpha_{n}}\in{\rm Str}(\mathcal{S}_{\alpha_{n-1}}),$ then
$\mathcal{S}_{\alpha_n}=\mathcal{S}_{\alpha_n-1}$ and $\zeta_{\alpha_n+}\in{\rm
Str},$ $\zeta_{\alpha_n-}\in{\rm Ctr}.$

Let $\zeta_{\alpha_{n}}\in{\rm Ctr}(\mathcal{S}_{\alpha_{n-1}}),$ then
$\mathcal{S}_{\alpha_{n}}=\zeta_{\alpha_{n}}(\mathcal{S}_{\alpha_{n-1}})$ and
$\zeta_{\alpha_{n+}}\in{\rm Ctr},$ $\zeta_{\alpha_n-}\in{\rm Str}.\square$

\begin{pred}\label{sign}
For any $n$ and $k$ $\mathcal{S}_{i_1\dots i_{k-1}+-i_{k+2}\dots
i_n}=\mathcal{S}_{i_1\dots i_{k-1}-+i_{k+2}\dots i_n}.$
\end{pred}

\begin{figure}[htb]
\centering \unitlength=0.50mm \special{em:linewidth 0.4pt}
\linethickness{0.4pt} \footnotesize \unitlength=0.70mm \special{em:linewidth
0.4pt} \linethickness{0.4pt}
\input{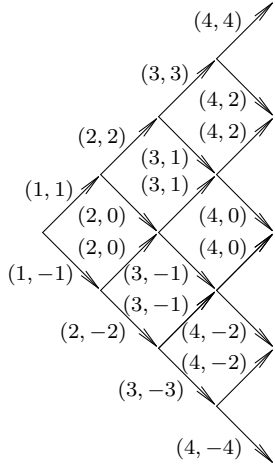}
\caption{\small Graph of simple self-intersections.}\label{si}
\end{figure}

\prf Accordingly to inductive definition
\[\zeta_{i_1\dots i_{k-1}+}:\ \mathcal{S}_{i_1\dots i_{k-1}}
\to\zeta(\mathcal{S}_{i_1\dots i_{k-1}});\ \ \zeta_{i_1\dots i_{k-1}-}:\
\mathcal{S}_{i_1\dots i_{k-1}}\to \zeta^{-1}(\mathcal{S}_{i_1\dots i_{k-1}}).
\]
Then
\[
\mathcal{S}_{i_1\dots i_{k-1}+}=\mathcal{S}_{i_1\dots i_{k-1}}
\cap\zeta(\mathcal{S}_{i_1\dots i_{k-1}}),\ \ \mathcal{S}_{i_1\dots
i_{k-1}-}=\mathcal{S}_{i_1\dots i_{k-1}} \cap\zeta^{-1}(\mathcal{S}_{i_1\dots
i_{k-1}}).\]
Similarly
\[\zeta_{i_1\dots i_{k-1}+-}:\ \mathcal{S}_{i_1\dots i_{k-1}+}\to
\zeta^{-1}(\mathcal{S}_{i_1\dots i_{k-1}+});\ \  \zeta_{i_1\dots i_{k-1}-+}:\
\mathcal{S}_{i_1\dots i_{k-1}-}\to\zeta(\mathcal{S}_{i_1\dots i_{k-1}-}).\]
Finally, we check
\[
\mathcal{S}_{i_1\dots i_{k-1}+-}=\mathcal{S}_{i_1\dots
i_{k-1}+}\cap\zeta^{-1}(\mathcal{S}_{i_1\dots
i_{k-1}+})=\zeta^{-1}(\mathcal{S}_{i_1\dots i_{k-1}})\cap\mathcal{S}_{i_1\dots
i_{k-1}}\cap\zeta(\mathcal{S}_{i_1\dots i_{k-1}})=\mathcal{S}_{i_1\dots
i_{k-1}-+}.\square
\]
So, all continuations of every deformation $\zeta$ can be depicted by the
following commutative branching partially ordered {\it graph $\Gamma$ of simple
self-intersections} (Fig.\ref{si}). Commutativity (convergence of arrows) is
guaranteed by proposition \ref{sign}. Notation $(n,s),$ which is shortening of
$\zeta_{(n,s)},$ includes $n$
--- {\it order of continuation} of $\zeta$ (length of binary code $\alpha_n$)
and $s$ ---  {\it signature of continuation} --- difference between number of
$+$ and $-$ within binary code $\alpha_n.$ Correctness and unambigiousity of
such notations is again guaranteed by proposition \ref{sign}. If some arrow
$(n_0,s_0)$ belongs to the ${\rm Simp},$ then all following arrows $(n,s)$ with
$n>n_0,\ s_0-(n-n_0)<s<s_0+(n-n_0)$ are the simplest accordingly to the
proposition \ref{stat}.
So, for every oriented path of graph $\Gamma$ there are two possible
alternatives: either on some step $(n_0,s_0)$ it become simplest, or it can be
infinitely prolonged as nonsimplest. In this last case we'll call  order of
self-intersection of $\zeta$  {\it infinite}. If any path of $\Gamma$ become
simplest on some step, we say that order of self-intersection of $\zeta$ is
{\it finite.} Then we can define type and order of this finite
self-intersection, specifying order and type of continued deformation, from
which the simplest types begin. Lets consider some examples.

\begin{figure}[htb]
\centering \unitlength=0.50mm \special{em:linewidth 0.4pt}
\linethickness{0.4pt} \footnotesize \unitlength=0.70mm \special{em:linewidth
0.4pt} \linethickness{0.4pt}
\input{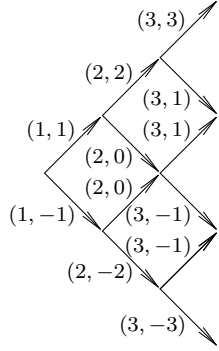}
\caption{\small Shift of square. Order and type of the graph --- $(3,{\rm
Par})$}\label{square}
\end{figure}

1. Consider parallel shift of square on $\mathbb{R}^2$ along diagonal on its
$1/3$ part. The deformation and its graph of self-intersection are shown in
Fig.\ref{square}

Beginning with $n=3$ the graph is stabilized and all $\zeta_{\alpha_3}$ belong
to the type ${\rm Par}.$ So the type of the graph is $(3,{\rm Par}).$

\begin{figure}[htb]
\footnotesize \centering \unitlength=0.50mm \special{em:linewidth 0.4pt}
\linethickness{0.4pt}
\input{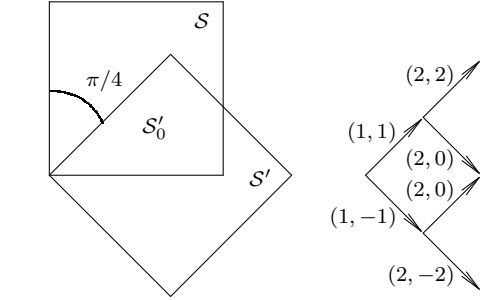}
\caption{\small Rotation of square. Order and type of the graph --- $(3,{\rm
Sl})$.}\label{rot}
\end{figure}

2. Lets consider rotation of square on $\mathbb{R}^2$ by angle $\pi/4$ around
one of its  vertexes (Fig.\ref{rot}).

Beginning with $n=2$ the graph is stabilized and all $\zeta_{\alpha_n},\ n\ge
2$ belong to the type ${\rm Sl}(O).$ The type of the graph is $(2,{\rm
Sl}(O)).$ Sliding set is center of rotation $O$.

\begin{figure}[htb]
\footnotesize \centering \unitlength=0.50mm \special{em:linewidth 0.4pt}
\linethickness{0.4pt}
\input{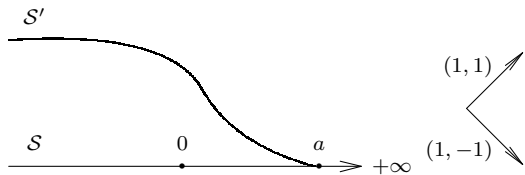}
\caption{\small Bending and shift of $\mathbb{R}^1.$ The type of the graph ---
$(1,{\rm Ctr},\ {\rm Str})$.}\label{split}
\end{figure}

3. Consider deformation of $\mathbb{R}^1$ in $\mathbb{R}^2,$ such that
$\mathcal{S}'$ is obtained from  $\mathcal{S}=\mathbb{R}^1$ by  bending
$\mathbb{R}^1$ at the point $0$ and by following constant shift  of obtained
curve on vector $(a,0)$ (along $\mathbb{R}^1$) (Fig.\ref{split}).
Easily to see, that under $n=1$ graph of self-intersection is stabilized.
Namely, $\zeta_+\in{\rm Ctr},\ \zeta_-\in{\rm Str}.$ So, its type is $(1, {\rm
Ctr},\ {\rm Str}).$

We have consider the case of simple self-in\-ter\-sec\-ti\-ons, when every
continuation $\mathcal{S}_{(n,s)}$ is connected. If it is not the case, we need
to introduce one additional index $\gamma_{(n,s)},$ numbering connected
components of $\mathcal{S}_{(n,s)}$ for every pair $(n,s):$
\[
\mathcal{S}_{(n,s)}=\bigcup\limits_{\gamma_{(n,s)}}\mathcal{S}_{\gamma_{(n,s)}}.
\]
Graph of self-intersection will become more complicated: it acquires additional
branching (say in third dimension) due to the possible topological branching of
continuation $\mathcal{S}_{(n,s)}.$ However, notions of finite and infinite
order of self-intersection remains valid and specifyings of finite order and
type of self-intersections are well defined.

The more detailed (but more complicated) classification of self-intersections
involves analysis of intersection matrix
$\mathcal{I}(\zeta_{(n_1,s_1)},\zeta_{(n_2,s_2)}).$ We don't touch this
possibility in the present paper.

Lets briefly outline the role of $\mathcal{I}(\zeta_1,\zeta_2).$ Firstly, we
observe, that for every $\zeta_1:\ \mathcal{S}_1\to\mathcal{S}_2$ and for every
$Y^{+}_{\zeta_1}\ni\zeta_2:\ \mathcal{S}_2\to\mathcal{S}_3$ or
$Y^{-}_{\zeta_1}\ni\zeta_2:\ \mathcal{S}_0\to\mathcal{S}_1$ we have:
\[
\mathcal{I}(\zeta_1,\zeta_2)=\left(\begin{array}{cc}
\mathcal{S}_1\cap\mathcal{S}_2& \mathcal{S}_1\cap\mathcal{S}_3\\ \mathcal{S}_2
&\mathcal{S}_2\cap\mathcal{S}_3
 \end{array}\right)
\ \ \mbox{\rm or}\ \ \mathcal{I}(\zeta_1,\zeta_2)=\left(\begin{array}{cc}
\mathcal{S}_1\cap\mathcal{S}_0& \mathcal{S}_1\\ \mathcal{S}_2\cap\mathcal{S}_0
&\mathcal{S}_2\cap\mathcal{S}_1
 \end{array}\right)
\]
respectively. It is naturally to call such class of intersection matrices and
deformations {\it consequent.}

Let we have the pair of consequent deformations $\zeta_1:\
\mathcal{S}_1\to\mathcal{S}'_1$ and $\zeta_2:\ \mathcal{S}'_1\to\mathcal{S}_2.$

\begin{pred}
There is following relations between self-intersection  and intersection
matrices:
\begin{equation}\label{rel}
\mathcal{I}(\zeta_1,\zeta_1)\cdot\mathcal{I}(\zeta_2,\zeta_2)=
(\mathcal{S}_1\cup\mathcal{S}_2)\cap\mathcal{S}'_1\cdot\mathcal{I}(\zeta_1,\zeta_2),
\end{equation}
where boolean matrix multiplication is defined as usually (line $\times$
column) with the help of boolean operations.
\end{pred}
\prf The proposition can be checked directly.$\square$

Particularly, it is follows from (\ref{rel}), that, if $\zeta_1$ and $\zeta_2$
are both parallel (i.e.
$(\mathcal{S}_1\cup\mathcal{S}_2)\cap\mathcal{S}'_1=\varnothing)$, then
$\mathcal{I}(\zeta_1,\zeta_1)\cdot\mathcal{I}(\zeta_2,\zeta_2)=\varnothing.$

There is necessary and sufficient  matrix criteria for the situation, when two
parallel consequent deformations gives parallel composition.

\begin{pred}
Two consequent deformations $\zeta_1$ and $\zeta_2$ together with  their
composition $\zeta_2\circ\zeta_1$ are parallel, if and only if
\[
\mathcal{I}(\zeta_1,\zeta_2)=\left(\begin{array}{cc} \varnothing &
\varnothing\\ \mathcal{S}'_1&\varnothing
 \end{array}\right)
\]
\end{pred}

\prf Proposition is checked  directly in both directions.$\square$

In case of more general situation we have

\begin{pred}
Two consequent deformations give parallel composition, if and only if
\[
\mathcal{I}(\zeta_1,\zeta_2)=\mathcal{S}'_1\left(\begin{array}{cc}
\mathcal{S}_1 & \varnothing\\ \mathcal{M}&\mathcal{S}_2
 \end{array}\right)
\]
\end{pred}

\prf Proposition is checked  directly in both directions.$\square$

At the end of the subsection we introduce some another special deformations.
We'll say, that $\zeta:\ \mathcal{S}\to\mathcal{S}'$ is {\it deformation with
invariant (fixed) set $\mathcal{S}^{\rm i}$ $(\mathcal{S}^{\rm f}),$} if
$\zeta|_{\mathcal{S}^{\rm i}}\in{\rm Sl}(\mathcal{S}^{\rm i})$
$(\zeta|_{\tilde\mathcal{S}}\in{\rm Sl}(\tilde\mathcal{S})$ for any
$\tilde\mathcal{S}\subseteq\mathcal{S}^{\rm f})$.

\subsection{Homotopies, histories and proper deformations}\label{hom}

Lets consider the set $\proj{H}{\mathcal{E}},$ consisting of homotopic classes
of embeddings $\mathcal{E}.$ Here we define {\it strong smooth homotopy} of
embedding $\iota\in\mathcal{E}$ as smooth mapping $F:\
\tm{\mathcal{B}}{I}\to\mathcal{M},$ where $I=[0,1],$ such, that
$F(\mathcal{B},0)=\iota$ and $F(\mathcal{B},t)\equiv
F_t(\mathcal{B})\in\mathcal{E}$ for every $t\in I.$ The two embeddings $\iota$
and $\iota'$ are said to be homotopic: $\hmt{\iota}{\iota'}$, if there exist
strong homotopy $F,$ such that $F_0(\mathcal{B})=\iota,$
$F_1(\mathcal{B})=\iota'.$ Homotopy relation is equivalence on $\mathcal{E}$
and $\pi_H(\mathcal{E}) \equiv\mathcal{E}/\hmt{}{}$ \cite{greenberg}.

Lets define strong homotopic equivalence on  $\tm{\mathcal{E}}{\mathcal{E}}.$
We'll say, that $\hmt{\de{1}{2}}{\depr{1}{2}},$ if simultaneously
$\hmt{\iota_1}{\iota'_1}$ and  $\hmt{\iota_2}{\iota'_2}.$ Obviously, the set of
classes of the strong homotopic equivalence
$\pi_H(\tm{\mathcal{E}}{\mathcal{E}})=\tm{\pi_H(\mathcal{E})}{\pi_H(\mathcal{E})}.$

Now we are able to define some special kinds of deformations in $\df,$ using
the homotopy relation. Lets consider the set $\pi^{-1}_H({\rm
Dg}(\tm{\proj{H}{\mathcal{E}}}{\proj{H}{\mathcal{E}}})),$ i.e. set of pair of
homotopic embeddings. The set, after factorization by $\pi_D$ becomes the
subset $\df_0\subseteq\df,$ which we'll call {\it proper deformations.}
With\-in the classical (nonquantum) $d$-structures it is naturally to restrict
ourself only by this type of deformations. Obviously,
 $\df_0$
---  subpseudogroup  of  $\df.$

For every  $\zeta\in\df_0$ by its definition there exists some {\it history}
--- strong homotopy $F_{(\zeta)t},$ such that $F_{(\zeta)0}={\rm
Dom}\,\zeta=\mathcal{S},$ $F_{(\zeta)1}={\rm
Im}\,\zeta\equiv\zeta(\mathcal{S})=\mathcal{S}'.$ The set of all histories of
the deformation $\zeta$ we'll denote $\Hs{\zeta}$ and call {\it class of
histories} of $\zeta$. It is easily to see, that pseudogroup structure on $\df$
induces composition law for histories: for every $\zeta_1,\zeta_2,\zeta_3,$
such that $\zeta_3=\zeta_2\circ\zeta_1,$ we put
\[\hs{\zeta_3}{}=\hs{\zeta_2\circ\zeta_1}{}\equiv\hs{\zeta_2}{}\circ\hs{\zeta_1}{},\]
where last equation means standard composition of homotopies \cite{greenberg}.
Similarly, we can define multiplication of classes
$\Hs{\zeta_2}\circ\Hs{\zeta_1}=\tm{\Hs{\zeta_2}}{\Hs{\zeta_1}}\subset\Hs{\zeta_3},$
consisting of all possible compositions of histories from $\Hs{\zeta_1}$ and
$\Hs{\zeta_2}.$

Every $\zeta\in\df_0$ can be classified by the methods of previous section.
We'll say, that history $F_{(\zeta)t}$ has type $\zeta$ {\it in a strong
sense,} if $F_{(\zeta)t}$ has the same type as $\zeta$  on a whole $I.$ Now we
can introduce {\it the simplest proper deformations} as collection
\[
{\rm Simp}_0\equiv{\rm Simp}\cap\df_0
\]
with histories of corresponding types in the strong sense. Also, we introduce
notions of {\it strongly invariant (fixed)} subset $\mathcal{S}^{\rm
i}\subseteq\mathcal{S}=\iota(\mathcal{B})$ ($\mathcal{S}^{\rm
f}\subseteq\mathcal{S}=\iota(\mathcal{B})$) relatively  $F_{(\zeta)t},$ if
\[
F_{(\zeta)t}(\mathcal{S}^{\rm i})=\mathcal{S}^{\rm i} \ \ (F_{(\zeta)t}(s)=s\
\mbox{\rm for all }\ s\in\mathcal{S}^{\rm f})\]  for all  $t\in I.$

\subsection{Vector fields,  motions and  generalized Killing
equations}\label{vector}

Lets consider some proper deformation \[\zeta\in\df_0:\
\mathcal{S}=\iota(\mathcal{B})\stackrel{\zeta}{\to}\mathcal{S'}=\iota'(\mathcal{B})\]
and let $\hs{\zeta}{}$ will be its some  history. Consider the set
$\mathcal{M}\supseteq\mathcal{P}_{\hs{\zeta}{}}=\cup_{t\in
I}\hs{\zeta}{t}(\mathcal{B})\equiv\cup_{t\in I}\mathcal{S}_t.$ It can be
treated as image of  smooth mapping of the smooth manifold
$I\times\mathcal{B}\to \mathcal{M},$ which, generally speaking, is not
submanifold and even not immersion in $\mathcal{M}.$ We'll call it {\it trace
of history} $F_{(\zeta)t}$ in $\mathcal{M}.$ Its boundary
$\partial\mathcal{P}_{F_{(\zeta)}}$ is $\mathcal{S}\cup\mathcal{S'}\cup_{t\in I
}F_t(\partial\mathcal{B}).$ Let $\widetilde{d/dt}$ ---  uniquely determined
horizontal vector field on $\tm{\mathcal{B}}{I},$ i.e. such that
$d\pi_1(\widetilde{d/dt})=0,$ $d\pi_2(\widetilde{d/dt})=d/dt,$ where
$\pi_1,\pi_2$ --- projections of $\tm{\mathcal{B}}{I}$ onto $\mathcal{B}$ and
$I$ respectively. The trace $\mathcal{P}_{\hs{\zeta}{}}$ is composed of an
integral lines $\{\hs{\zeta}{t}(b)\}_{b\in\mathcal{B}}$ of the vector field
$\tau=d\hs{\zeta}{}(\widetilde{d/dt}),$ defined on $\mathcal{P}_{F_{(\zeta)}}.$
The family of embeddings $\{\hs{\zeta}{t}(\mathcal{B})\}_{t\in I}$ induces the
family of deformation forms $\{\Delta^t_\mathcal{B}\}_{t\in I}$ by the
following rule:
\[
\Delta^t_{\mathcal{B}}=(d\hs{\zeta}{t})^\ast\Theta-(d\hs{\zeta}{0})^\ast\Theta.
\]
We'll say, that the history $\hs{\zeta}{t}$ is {\it motion}\footnote{Of course,
one can consider  not only proper motion. The generalization is obvious and we
don't touch it in present paper.} of the $d$-body $\mathcal{B}$ in
$\mathcal{M},$ if $\Delta^t_{\mathcal{B}}=0$ for any $t\in I$ or, in other
words, if the image $(d\hs{\zeta}{t})^\ast\Theta$ is constant on $I$. This
notion generalizes a concept of absolutely rigid solids in classical mechanics.

\begin{pred}\label{motlie}
History $\hs{\zeta}{t}$ is motion, if and only if
\begin{equation}\label{killing}
\left.\pounds_\tau\Theta\right|_{\mathcal{P}_{F_{(\zeta)}}}=0,
\end{equation}
where $\pounds_\tau$ --- Lie derivative along the vector field
$\tau=d\hs{\zeta}{}(\widetilde{d/dt})$ on $\mathcal{P}_{\hs{\zeta}{}}.$
\end{pred}

\prf\ Lets calculate the derivative \[
\frac{d}{dt}\left\{(d\hs{\zeta}{t})^{\ast}\Theta_{\mathcal{S}_t}\right\}\stackrel{1}{=}\lim_{s\to
0}\frac{(d\hs{\zeta}{t+s})^{\ast}\Theta_{\mathcal{S}_{t+s}}-(d\hs{\zeta}{t})^\ast\Theta
_{\mathcal{S}_t}}{s}\]
\[\stackrel{2}{=}(d\hs{\zeta}{t})^\ast\lim_{s\to
0}\frac{((d\hs{\zeta}{t})^\ast)^{-1}\circ(d\hs{\zeta}{t+s})^{\ast}\Theta_{\mathcal{S}_{t+s}}-
\Theta_{\mathcal{S}_t}}{s}\]
\[\stackrel{3}{=}(d\hs{\zeta}{t})^\ast\lim_{s\to 0}
\frac{(d(\hs{\zeta}{t+s}\circ
(F_{(\zeta)t})^{-1}))^\ast\Theta_{\mathcal{S}_{t+s}}-\Theta_{\mathcal{S}_t}}{s}\]
\[\stackrel{4}{=} (d\hs{\zeta}{t})^\ast \pounds_\tau\Theta_{\mathcal{S}_t}.\] The
eq. "1"
--- definition of derivative,  in "2" we have used independence of
$(dF_{(\zeta)t})^\ast$ on $s,$ in "3" --- property (\ref{codiff}) and identity
$((d\alpha)^\ast)^{-1}=(d(\alpha^{-1}))^\ast$ for any diffeomorphism $\alpha,$
and in "4" - definition of Lie derivative (note, that the mapping $F_{t+s}\circ
F_t^{-1}$ maps $\mathcal{S}_t\to\mathcal{S}_{t+s}$). Since
$(dF_{{(\zeta)}t})^\ast$ is nondegenerate under every fixed $t$, then
proposition is proved.$\square$

 The equations (\ref{killing}) we'll call {\it generalized
Killing equations,} and  $\tau$ --- {\it generalized Killing vector field.}

\subsection{$d$-coverings of $d$-manifolds}\label{cover}

Let $\mathcal{S}=\iota(\mathcal{B})$ will be some fixed deformant  and let
$\mt{\mathcal{S}}$ --- set of all its  possible  motions in $\mathcal{M}.$
Easily to see, that the motions define equivalence $\stackrel{M}{\sim}$ on
$\mathcal{E}$: we'll call the two embeddings $\iota$ and $\iota'$
$M-$equivalent: $\iota\stackrel{M}{\sim}\iota',$ if there exist history
$\hs{\zeta}{}\in\Hs{\zeta},$ where $\zeta:\ \mathcal{S}\to\mathcal{S}',$ such
that $\hs{\zeta}{}\in\mt{\mathcal{S}}.$ Obviously, the equivalence
$\stackrel{M}{\sim}$ is more weak then $\stackrel{H}{\sim}$ and, generally
speaking, the set $\pi_H^{-1}[\iota]_H=\cup_\alpha\pi_M^{-1}[\iota_\alpha]_M,$
where $\{\iota_\alpha\}$
--- some set of all pair-wisely $\stackrel{M}{\sim}$-nonequivalent elements from
$\pi_H^{-1}[\iota]_H,$ and
$\pi_M^{-1}[\iota_\alpha]_M\cap\pi_M^{-1}[\iota_\beta]_M=\varnothing$ for all
$\alpha\neq\beta.$ We'll call $\pi_M^{-1}[\iota_\alpha]_M$ --- {\it
$\alpha$-component} of $\pi_H^{-1}[\iota]_H,$ and its image
\[\mathcal{R}({\mathcal{S}_\alpha})\equiv\bigcup\limits_{F_{(\zeta)}\in
\mt{\mathcal{S}_\alpha}}\mathcal{P}_{F_{(\zeta)}}\] {\it rigidity
$\alpha-$component of the manifold $\mathcal{M}$ relatively to the embedding
$\iota.$} Here $\mathcal{S}_\alpha=\iota_\alpha(\mathcal{B}).$ The family
$\{\mathcal{R}(\mathcal{S}_\alpha)\}$ forms  some covering of $\mathcal{M}$:
\[\mathcal{M}=\bigcup\limits_{\alpha}\mathcal{R}(\mathcal{S}_\alpha),\] which
we'll call {\it deformational $(\mathcal{B},\Theta,h)-$co\-ve\-ring of the
manifold $\mathcal{M}$} or, more shortly, {\it $d$-covering,} where
$\proj{H}{\mathcal{E}}\ni h=\pi_H(\iota)$,

Within the classical dynamical $d$-structures, which will be considered in the
next sections, it is naturally to use as configuration space of deformant not
$\pi_H^{-1}[\iota]_H,$ but its factor:
\[\{\pi_H^{-1}[\iota]_H/\stackrel{M}{\sim}\}\equiv\{\pi_{M}\circ\pi_H^{-1}
[\iota]_H)\} \cong\{\mathcal{R}(\mathcal{S}_\alpha)\},\] that reflects the
deformational indistinguishability  of those configurations, that are connected
by  some motion. It will be  automatically provided in second half of the paper
by formulation of physical action $\mathfrak{F}$ in terms of $\Delta:$
$\mathfrak{F}=\mathfrak{F}[\Delta],$ so that
$\delta\mathfrak{F}\sim\delta\Delta$
--- vanishes on motions.

We'll call the manifold $\mathcal{M}$ {\it deformationally trivial} relatively
to its $(\mathcal{B},\Theta,h)-d$-co\-ve\-ring, if $\pi_M$
--- constant mapping and {\it deformationally discrete,}  if $\pi_M$
--- identical mapping. The manifold $\mathcal{M}$ will be called
{\it deformationally homogeneous ($d$-ho\-mo\-ge\-ne\-o\-us),} if
\begin{equation}\label{homog}
\mathcal{R}(\mathcal{S}_\alpha)=\mathcal{M}
\end{equation}
for some  $\alpha$ and  {\it completely deformationally homogeneous,} if
(\ref{homog}) is satisfied for all $\alpha.$

Deformationally trivial manifolds have no significance from the view point of
deformational structure theory by the following
\begin{pred}
Any deformationally trivial manifold has:

1) constant $d$-metric, if $\Theta\in\Omega^0;$

2) zero $d$-metric, if $\Theta\in\Omega^{\otimes p},\ p\neq0.$
\end{pred}

\prf For any of the cases deformational triviality by the proposition 8 means
$\pounds_\tau\Theta=0$ for all smooth vector fields $\tau.$  In case $p=0$ we
have $\pounds_\tau\Theta=\tau\Theta=0$ and then, in any coordinate system
$\{x^A\}$ on $\mathcal{M}$, taking consequently $\tau=\partial_A,\ A=1,\dots,n$
we obtain $\partial_A\Theta=0,\ A=1,\dots,n\Rightarrow\Theta={\rm const}.$

In case $p\neq0$ we have in coordinates:
\begin{equation}\label{Lie}
(\pounds_\tau\Theta)_{A_1\dots A_p}=\tau^B\partial_B\Theta_{A_1\dots A_p}
+\sum\limits_{j=1}^p (\partial_{A_j}\tau^B)\Theta_{A_1\dots B\dots A_p}=0
\end{equation}

Take as  previously $\tau=\partial_A,\ A=1,\dots,n$ consequently and obtain
that $\Theta$ is constant form (so the first term in (\ref{Lie}) vanishes).
Since coordinate system is arbitrary, we conclude, that $\Theta\equiv0.\square$

 Riemannian manifold with general metrics $g$ is an example of
deformationally discrete manifold. Euclidean space $\mathbb{E}^n$ is completely
deformationally homogeneous relatively any
$(\mathcal{B},\eta,h)-$decomposition, whe\-re $\eta$ --- Euclidean metric,
$\mathcal{B}$ --- arbitrary $d$-body, $h$
--- arbitrary element $\proj{H}{\mathcal{E}}.$
As an example of deformationally homogeneous, but not completely
deformationally homogeneous manifold consider the following situation. Let
$\mathcal{M}=\overline{D^2_{2r}}(0)\setminus D^2_r(0)$ --- closed ring on 2D
Euclidean plane (as usually, $D^n_r(a)$ --- $n-$dimensional disk with radius
$r$ and center $a,$ bar above letter --- topological closure), $\Theta=\eta$
--- 2D Euclidean metrics, $\mathcal{B}=S^1,$ $\iota(S^1)=S^1_{R}\subset\mathcal{M}$ ---
circle with radius $R$ and $\pi_H(\iota)=1$ (for the considered case
$\pi_H(\mathcal{E})\equiv\pi_1(\mathcal{M})$
--- fundamental group of $\mathcal{M},$ isomorphic $\mathbb{Z}.$) Then, in case
$R<3r/2,$  $\mathcal{R}(S^1_R)=\overline{D^2_{2R-r}}(0)\setminus
D^2_r(0)\neq\mathcal{M}$ and only in case  $R=3r/2,$
$\mathcal{R}(S^1_{3r/2})=\mathcal{M}.$

Now we formulate two propositions and give an example, all illustrating
relation of a free deformational structure theory with isometries of Riemannian
spaces.

Let ${\rm St}(v)\equiv\{m\in\mathcal{M}\,|\ \psi_t(m)=m\}$ will be the set of
all stationary  points of  one-parametric group $\psi_t$, generated by some
smooth vector field $v\in T\mathcal{M}.$

\begin{pred}\ If manifold $\mathcal{M}$ admits isometry of $d$-metrics,
i.e. if there exists vector field $v\in T\mathcal{M},$ such that
$\pounds_v\Theta=0,$ then $\forall\ \mathcal{S}$ such that
$\mathcal{S}\not\subseteq{\rm St}(v),$ there exists nonidentical motion
$\{\left.\psi_t\right|_{\mathcal{S}}\}\in$ $\mt{\mathcal{S}}$ and, by the fact,
$\mathcal{M}$ is not deformationally discrete.
\end{pred}

\prf The proposition immediately follows from the relation: $\pounds_v\Theta=0\
\Rightarrow\ \pounds_{\tilde v}\left.\Theta\right|_{\mathcal{P}_F}=0,$ where
$\mathcal{S}_t=F_t(\mathcal{B})\equiv \psi_t|_{\mathcal{S}},$ $\tilde v\equiv
v|_{\mathcal{P}_F}.\square$

\begin{pred}
If manifold $\mathcal{M}$ admits $r-$parametric isometry group $\mathcal{G}$,
generated by  vector fields $\left\{v_1,\dots\right.$ $\left.,v_r\right\},$
such that $\pounds_{v_i}\Theta=0,\ i=1,\dots,r,$ that acts on $\mathcal{M}$

1) transitively, then $\mathcal{M}$ --- completely deformationally homogeneous
(relatively any decomposition);

2) intransitively, and if also $\mathcal{S}\cap{\rm Orb}\, \mathcal{G}$
--- connected for some orbit ${\rm Orb}\,\mathcal{G}$, then

a) ${\rm Orb}\,\mathcal{G}$ --- completely deformationally homogeneous
relatively its $(\iota^{-1}(\mathcal{S}\cap{\rm
Orb}\,\mathcal{G}),\left.\Theta\right|_{{\rm Orb}\,\mathcal{G}},
h=\pi_{H}(\iota))$-decomposition.

b) $\mathcal{G}\le\mt{\mathcal{S}\cap{\rm Orb \mathcal{G}}}.$

Here, as usually, $\mathcal{S}=\iota(\mathcal{B}).$
\end{pred}

\prf 1) Taking any $\mathcal{S}$ and acting by $\mathcal{G},$ we get (by the
transitivity property):
\[\bigcup\limits_{g\in\mathcal{G}}g(\mathcal{S})=\mathcal{M}.\]

2) (a) follows from transitivity property of $\mathcal{G}$ on ${\rm
Orb}\,\mathcal{G}.$ (b)  is obvious.$\square$

 Particularly, if $\mathcal{S}=\iota(\mathcal{B})={\rm
Orb}\,\mathcal{G},$ then \[\mt{\mathcal{S}}\cap{\rm
Sl}(\mathcal{S})\neq\varnothing\] defines the group of {\it rigid proper
slidings.}

So, if $\Theta$ --- Riemannian (or any other $d$-) metric on  $\mathcal{M}$ and
$\mathcal{M}$ admits isometry,  then nontrivial motions of $d$-objects always
exist. The following example shows, that inverse is not always valid.

Let $\mathcal{M}=\mathbb{R}^2$ with cartesian coordinate system $\{x_1,x_2\},$
$\mathcal{B}=I=[0,1]\in\mathbb{R},$
$\Omega^1(\mathbb{R}^2)\ni\Theta=(x^1x^2+\coth x^2)dx^1.$ Let
$\iota(\mathcal{B})\equiv\mathcal{S}=\{0\le x^1\le 1,\,x^2=0\}.$ By the fact,
that $\left.\Theta\right|_{x^2=0}=dx^1={\rm const},$ it is easily to see that
the set of homotopies \[\{F_t: \mathcal{S}\to\mathcal{S}_t=(x_1+t,0),\ 0\le
x_1\le1,\ -\infty<t<\infty\},\] (they are simple rigid translations of units
interval along axe $x^1$) lies in $\mt{\mathcal{S}}.$ Moreover,
$\mathcal{P}_{F_t}=\mathbb{R}^1=\{(x^1,0)\}.$ The related vector field $\tilde
v(t,x^1),$ along which
 $\pounds_{\tilde v}\left.\Theta\right|_{\mathcal{P}_F}=0$ is simply $\partial/\partial x^1.$
It is easily to show, that $\tilde v$ does'nt admit smooth continuation $ v$
from $\mathcal{P}_{F}$ on a whole $\mathbb{R}^2.$ Really, Killing equations
$\pounds_{v}\Theta=0$ for this case (under restriction $\left.
v\right|_{\mathcal{P}_{F}}=\partial/\partial
 x^1$) ultimately give:
 \[v^2=-\frac{x^2}{x^1+\sinh x^2}.\] The component has singularity
 on line $x^1=-\sinh x^2,$ which cross any neighborhood of
 $\mathcal{P}_{F}$ in $\mathbb{R}^2.$

We'll call  the set
\[
\mt{\mathcal{B}}=\bigcup\limits_{\iota\in\mathcal{E}}\mt{\iota(\mathcal{B})}\]
{\it motions of $d$-body in $\mathcal{M}.$} Obviously,
$\mt{\mathcal{B}}<\df_0.$

\subsection{Conformal motions}\label{conf}

Similarly to Riemannian geometry we also define more general (then motions)
histories --- {\it conformal motions.} Infinitesimally, they are defined by the
equation:
\begin{equation}\label{cmot}
\frac{d}{dt}\left\{(dF_{(\zeta)t})^\ast\Theta\right\}=\varphi\cdot(dF_{(\zeta)t})^\ast\Theta,
\end{equation}
where $\varphi:\ \mathcal{B}\times I\to\mathbb{R}$ --- some scalar function.
Using calculations similar to proof of proposition \ref{motlie}, it is easily
to show, that (\ref{cmot}) is equivalent to the following {\it generalized
conformal Killing equations:}
\[
\pounds_\tau\left(\Theta|_{\mathcal{P}_F}\right)=\varphi\Theta|_{\mathcal{P}_F},
\]
where $\tau$ --- {\it generalized conformal vector field.}
 We'll denote all possible histories with initial embedding
$\mathcal{S}$, satisfying (\ref{cmot}), $\mathrm{C}\mt{\mathcal{S}}_{\varphi}$
and set of  such histories for all $\mathcal{S}$
$\mathrm{C}\mt{\mathcal{B}}_{\varphi}.$ Obviously, that
$\mt{\mathcal{B}}\le\mathrm{C}\mt{\mathcal{B}}_{\varphi}\le\df_0$ and
$\mt{\mathcal{B}}=\mathrm{C}\mt{\mathcal{B}}_0.$ Similarly to the case of
motions, we can define {\it conformal} deformational
$(\mathcal{B},\Theta,h)-$co\-ve\-ring of the manifold $\mathcal{M},$ and
conformal generalizations  of $d$-trivial, $d$-discrete and (completely)
$d$-homogeneous manifolds.

\subsection{$d$-substructures, compositions and polymetric $d$-structures}\label{mapp}

Lets define isomorphism between free $d$-structures. The two $d$-structures
\[ \dst{D}{}{B}{}{M}{}{E}{}{\Theta}\ \mbox{\rm and}\
\dst{D}{'}{B}{'}{M}{'}{E}{'}{\Theta'}\] will be called isomorphic, if there
exist diffeomorphisms \[\Psi:\ \mathcal{B}'\to\mathcal{B},\ \Phi:\
\mathcal{M}'\to\mathcal{M},\] such that
\[\Psi(\mathcal{B}')=\mathcal{B};\ \ \Phi(\mathcal{M}')=\mathcal{M};\ \ (d\Phi)^\ast\Theta=\Theta';\ \ \
\mathcal{E}'=\Phi^{-1}\circ\mathcal{E}\circ\Psi.\]

\begin{pred}
Isomorphic $d$-structures have isomorphic pseudogroups of deformations and
motions\footnote{As in case of groups we define {\it homomorphism between
pseudogroups} $\mathcal{A}_1$ and $\mathcal{A}_2$ as a mapping $\alpha:\
\mathcal{A}_1\to \mathcal{A}_2$, such that for every
$a_1,a_2,a_3\in\mathcal{A}_1$ connected by the relation $a_1\ast a_2=a_3$ takes
place relation for images $\alpha(a_1)\ast\alpha(a_2)=\alpha(a_3),$ where
$\ast$ in last expression --- pseudogroup multiplication in $\mathcal{A}_2.$
Also, we define {\it left} and {\it right kernels} of homomorphism for element
$a$ as the following subsets of $\mathcal{A}_1:$
\[\ker^a_L\alpha\equiv\{b\in\mathcal{A}_1\,|\, (b,a)\in \mathcal{U}_1,\ \alpha(b)=e^-_{\alpha(a)}\};\ \
\ker^a_R\alpha\equiv\{b\in\mathcal{A}_1\,|\, (a,b)\in \mathcal{U}_1,\
\alpha(b)=e^+_{\alpha(a)}).\] In case $\ker^a_{L,R}\alpha=e^{\mp}_a\ \forall
a\in\mathcal{A},\ \alpha$ --- isomorphism of pseudogroups.}.
\end{pred}

\prf\ Isomorphism \[\varsigma:\
\vardf{\mathcal{M}}{\mathcal{B}}\to\vardf{\mathcal{M}'}{\mathcal{B}'}\] is
given by the relations:
\[
\varsigma(\zeta)=\Phi^{-1}\circ\zeta\circ\Phi,\
\mathcal{S}'_i=\Phi^{-1}(\mathcal{S}_i), \ i=1,2
\]
for every $\df\ni\zeta:\ \mathcal{S}_1\to\mathcal{S}_2.$

If $F_t\in\mt{\mathcal{B}},$ then $F'_t=\Phi^{-1}\circ
F_t\circ\Phi\in\varmt{\mathcal{M}'}{\mathcal{B}'},$ since
\[
\frac{d}{dt}\left\{(dF'_t)^\ast\Theta'\right\}=\frac{d}{dt}\left\{(d(\Phi^{-1}\circ
F_t\circ\Phi))^\ast(d\Phi)^\ast\Theta\right\}=\frac{d}{dt}\left\{(d(F_t\circ\Phi))^\ast\Theta\right\}=
(d\Phi)^\ast\frac{d}{dt}\left\{(dF_t)^\ast\Theta.\right\}
\]
Since $(d\Phi)^\ast$ --- isomorphism, then also
\[
F'_t\in\varmt{\mathcal{M}'}{\mathcal{B}'}\Rightarrow F_t=\Phi\circ
F'_t\circ\Phi^{-1}\in\mt{\mathcal{B}},
\]
and so $\varmt{\mathcal{M}'}{\mathcal{B}'}\cong\mt{\mathcal{B}}.\square$

We'll call $d$-structure
$\mathfrak{D}'=\langle\mathcal{B}',\mathcal{M}',\mathcal{E}',\Theta'\rangle$
{\it $d$-substructure} of $\dst{D}{}{B}{}{M}{}{E}{}{\Theta},$ if
$\mathcal{B}'\subseteq \mathcal{B}$ is embedding of $\mathcal{B}'$ in
$\mathcal{B}$ or (and) $\mathcal{M}'\subseteq\mathcal{M}$ is embedding of
$\mathcal{M}'$ in $\mathcal{M}$ and $\Theta'=\Theta|_{\mathcal{M}'}$or (and)
$\mathcal{E}'\subseteq\mathcal{E}.$ In case "or"  some components of
$d-$structures may be identical. We shall denote this situation as
$\mathfrak{D}'\preceq_X\mathfrak{D},$ where $X$ shows restricted elements of
$\mathfrak{D},$ for example
$\mathfrak{D}'\preceq_\mathcal{B'}\mathfrak{\mathfrak{D}}.$

\begin{pred}
In case  $\mathfrak{D}'\preceq_{\mathcal{B}'}\mathfrak{D}$ there is
homomorphism $\alpha:\ \df\to\vardf{\mathcal{M}}{\mathcal{B}'}.$ In case
$\mathfrak{D}'\preceq_{\mathcal{M}'}\mathfrak{D}$ or
$\mathfrak{D}'\preceq_{\mathcal{E}'}\mathfrak{D},$
$\vardf{\mathcal{M}'}{\mathcal{B}}\le\df.$
\end{pred}

\prf In case $\mathfrak{D}'\preceq_{\mathcal{B}'}\mathfrak{D}$ homomorphism
$\alpha$  acts by the rule:
\[
\vardf{\mathcal{M}}{\mathcal{B}'}\ni\alpha(\zeta)=\alpha([\de{1}{2}]_D)
=[(\iota_1|_{\mathcal{B}'},\iota_2|_{\mathcal{B}'})]_{D'},\ \forall\zeta\in\df,
\]
where $D'$  means factorization by
$\Diff{\mathcal{B}}_{\mathcal{B}'}<\Diff{\mathcal{B}}$ with invariant
submanifold $\mathcal{B}'\subseteq\mathcal{B}.$ In case
$\mathfrak{D}'\preceq_{\mathcal{M}'}\mathfrak{D}$ it is obviously, that if
$\zeta'\in\vardf{\mathcal{M}'\subseteq\mathcal{M}}{\mathcal{B}},$ then
necessarily $\zeta'\in\df.$ Third case is obvious. One example of the case we
already have faced with: {\it proper substructure}
$\mathfrak{D}'\preceq_{\mathcal{E}'}\mathfrak{D}$ with $\df_0<\df.\square$

We'll say that free $d$-structure
$\mathfrak{D}=\langle\mathcal{B},\mathcal{M},\mathcal{E},\Theta\rangle$ is {\it
composition} of the free $d$-structures\\
$\mathfrak{D}_1=\langle\mathcal{B}_1,\mathcal{M}_1,\mathcal{E}_1,\Theta_1\rangle$
and
$\mathfrak{D}_2=\langle\mathcal{B}_2,\mathcal{M}_2,\mathcal{E}_2,\Theta_2\rangle$:
$\mathfrak{D}=\mathfrak{D}_1\times\mathfrak{D}_2,$ if
\[
\mathcal{B}=\tm{\mathcal{B}_1}{\mathcal{B}_2},\ \
\mathcal{M}=\tm{\mathcal{M}_1}{\mathcal{M}_2},\ \
\mathcal{E}=\tm{\mathcal{E}_1}{\mathcal{E}_2},\ \
\Theta=(d\pi_1)^\ast\Theta_1\otimes(d\pi_2)^\ast\Theta_2,
\]
where $\pi_1$ and $\pi_2$ --- projections of
$\tm{\mathcal{M}_1}{\mathcal{M}_2}$ onto multipliers.

\begin{pred}\label{twocomp}
For composite $d$-structure $\mathfrak{D}$
\[\df=\tm{\vardf{\mathcal{M}_1}{\mathcal{B}_1}}{\vardf{\mathcal{M}_2}{\mathcal{B}_2}},\
\ \mt{\mathcal{B}}=\tm{\mathrm{C}\varmt{\mathcal{M}_1}{\mathcal{B}_1}_\varphi}
{\mathrm{C}\varmt{\mathcal{M}_2}{\mathcal{B}_2}_{-\varphi}},\] where
$\varphi={\rm const}_{\mathcal{B}}.$
\end{pred}

\prf Note, that any pair
\[(\zeta_1,\zeta_2)\in\tm{\vardf{\mathcal{M}_1}{\mathcal{B}_1}}{\vardf{\mathcal{M}_2}{\mathcal{B}_2}}\]
defines unique deformation \[\zeta\in\df:\ \mathcal{S}\to\mathcal{S}',\] where
$\mathcal{S}=\tm{\mathcal{S}_1}{\mathcal{S}_2},\
\mathcal{S}'=\tm{\mathcal{S}'_1}{\mathcal{S}'_2}.$ Inversely, any $\zeta\in\df$
determines unique $\zeta_1=\pi_1\zeta$ and $\zeta_2=\pi_2\zeta.$ We only need
to restrict general diffeomorphisms $\Diff{\mathcal{B}}$ on its subgroup
$\tm{\Diff{\mathcal{B}_1}}{\Diff{\mathcal{B}_2}},$ conserving $\mathcal{B}_1$
and $\mathcal{B}_2$ in $\mathcal{B}$ and consistent with product structure of
$\mathcal{E},$ when define $\df$ as factor
$\tm{\mathcal{E}}{\mathcal{E}}/\stackrel{D}{\sim}.$

To clear out relation between motions pseudogroups, lets calculate the
derivative:
\[
\frac{d}{dt}\left\{(dF_t)^\ast\Theta\right\}=\frac{d}{dt}\left\{(dF_t)^\ast(d\pi_1)^\ast\Theta_1
\otimes (dF_t)^\ast(d\pi_2)^\ast\Theta_2\right\},
\]
where $F_t$ --- some history of some deformation $\zeta=(\zeta_1,\zeta_2)$ in
$\mathcal{M}$. Using composition property (\ref{codiff}),  relations $\pi_1
F_t=F_{1t},$ $\pi_2 F_t=F_{2t},$ where $F_{1t}\in\Hs{\zeta_1},$
$F_{2t}\in\Hs{\zeta_2}$ and Leibnitz rule we obtain:
\[
\frac{d}{dt}\left\{(dF_t)^\ast\Theta\right\}
=\frac{d}{dt}\left\{(dF_{1t})^\ast\Theta_1\right\}\otimes
(dF_{2t})^\ast\Theta_2+(dF_{1t})^\ast\Theta_1\otimes
\frac{d}{dt}\left\{(dF_{2t})^\ast\Theta_2\right\}.
\]
It is easily to see, that if
\[\frac{d}{dt}((dF_{1t})^\ast\Theta_1)=\varphi\cdot(dF_{1t})^\ast\Theta_1,\ \
\frac{d}{dt}((dF_{2t})^\ast\Theta_2)=-\varphi\cdot(dF_{2t})^\ast\Theta_2,\]
then previous equations are satisfied and so
$\tm{\mathrm{C}\varmt{\mathcal{M}_1}{\mathcal{B}_1}_\varphi}{\mathrm{C}\varmt{\mathcal{M}_2}{\mathcal{B}_2}_{-\varphi}}\subseteq\mt{\mathcal{B}}.$
Inversely, let $d/dt((dF_t)^\ast\Theta)=0.$ It means, that for any sets of
vector fields $\{u,v\}$ on $\mathcal{B}$: $u=\{u_1,\dots u_{p_1}\}\in
(T\mathcal{B}_1)^{\times p_1},$  $v=\{v_1,\dots v_{p_2}\}\in
(T\mathcal{B}_2)^{\times p_2},$ where $p_1={\rm deg}\,\Theta_1,$ $p_2={\rm
deg}\, \Theta_2,$ we have $d/dt((dF_t)^\ast\Theta)(u,v)=0.$ Lets denote
$(dF_{1t})^\ast\Theta_1(u)=f_1(u,t),\ (dF_{2t})^\ast\Theta_2(v)=f_2(v,t).$ Then
\[
d/dt((dF_t)^\ast\Theta)(u,v)=\dot f_1(u,t)f_2(v,t)+f_1(u,t)\dot f_2(v,t)
=\frac{d}{dt}(f_1f_2)=0,
\]
that gives $f_1f_2={\rm const}_I$ under any $u$ and $v.$ Then it follows, that
\[f_1(u,t)=\tilde f_1(u)\alpha(t),\ f_2(v,t)=\tilde f_2(v)/\alpha(t).\] Coming
back to codifferential, omitting arguments $u,v$ by its arbitrariness and
taking derivatives over $t$ we obtain
\[\frac{d}{dt}((dF_{1t})^\ast\Theta_1)=\varphi(t)(dF_{1t})^\ast\Theta_1;\ \
\frac{d}{dt}((dF_{2t})^\ast\Theta_2)=-\varphi(t)(dF_{2t})^\ast\Theta_2,\] with
$\varphi(t)=\alpha'/\alpha.$ We find that
$\mt{\mathcal{B}}\subseteq\tm{\mathrm{C}\varmt{\mathcal{M}_1}
{\mathcal{B}_1}_{\varphi}}{\mathrm{C}\varmt{\mathcal{M}_2}{\mathcal{B}_2}_{-\varphi}}$
with $\varphi=\varphi(t)$ and so, ultimately,
$\mt{\mathcal{B}}=\tm{\mathrm{C}\varmt{\mathcal{M}_1}{\mathcal{B}_1}_{\varphi}}
{\mathrm{C}\varmt{\mathcal{M}_2}{\mathcal{B}_2}_{-\varphi}}$ with
$\varphi=\varphi(t).\square$

There is direct generalization of proposition \ref{twocomp}.
\begin{pred}
For multicomponent $d$-structure
$\mathfrak{D}=\mathfrak{D}_1\times\dots\times\mathfrak{D}_n$
\[
\df=\prod\limits_{i=1}^{n}\vardf{\mathcal{M}_i}{\mathcal{B}_i};\ \
\mt{\mathcal{B}}=\prod\limits_{i=1}^{n}\varmt{\mathcal{M}_i}{\mathcal{B}_i}_{\varphi_i}
\]
with $\varphi_i=({\rm const}_\mathcal{B})_i$ and
$\sum\limits_{i=1}^n\varphi_i=0.$
\end{pred}

 Note, that in case
of composed $d$-structure $\mathfrak{D}=\mathfrak{D}_1\times\mathfrak{D}_2$
there is two independent $d$-metrics on
$\mathcal{M}=\tm{\mathcal{M}_1}{\mathcal{M}_2}$: $(d\pi_1)^\ast\Theta_1$ and
$(d\pi_2)^\ast\Theta_2,$ which we have used for constructing universal
$d$-metric on $\mathcal{M}$ with "good" properties. The similar situation
arises in a more general case, when $d$-manifold possess two (or more) metrics.
We'll call $d$-structure with the set of metrics $\{\Theta_\alpha\}$ on the
same $d$-manifold $\mathcal{M}$ {\it polymetric $d$-structure.} If we consider
all of $\Theta_\alpha$ as $d$-metrics, then every $\Theta_\alpha$ burns its own
pseudogroup of motions $\mt{\mathcal{B}}_\alpha.$ We can introduce partial
order on the set $\{\Theta_\alpha\}.$ Namely we define
$\Theta_\alpha\preceq\Theta_\beta,$ if
$\mt{\mathcal{B}}_\alpha\supseteq\mt{\mathcal{B}}_\beta.$ We'll say, that in
this case the metric $\Theta_\alpha$ is {\it weaker} then $\Theta_\beta.$
Obviously, that if $d$-metric $\Theta_\alpha$ is weaker then $\Theta_\beta,$
then $d$-covering induced by $\Theta_\beta$ is submitted to the $d$-covering,
induced by $\Theta_\alpha.$

Generally speaking, it is not necessary to consider  all metrics from
$\{\Theta_\alpha\}$ as $d$-metrics (see example in Sec.\ref{maxwell}). Some of
them can be used as $g$-metrics (see Sec.\ref{defdin}).

\subsection{Physical realizations of free\\ $d$-structures}\label{example1}

Any smooth form $\Theta\in\Omega^{\otimes p}$ defined on arbitrary manifold
$\mathcal{M}$ can be viewed as $d$-metric, if one specifies some $d$-body
$\mathcal{B}.$ So, any $\mathcal{M},$ supported by smooth form can be
transformed into some $d$-structures. In physical applications the most often
case is $\mathcal{B}=\mathcal{M}.$ In this case pseudogroup $\df$ become {\it
group of deformations of} $\mathcal{M},$ while $\mt{\mathcal{B}}$
--- its {\it subgroup of motions of $d$-metric $\Theta.$} Direct physical
realizations of  the such $d$-structures are following:
\begin{enumerate}
\item
 $\ \mathcal{M}=\mathcal{B}=\mathbb{E}^3$ --- 3D Euclidean space,  $\Theta=\eta$
--- Euclidean metrics. In this case we obtain kinematics of standard
elasticity theory (where $\Delta/2$ --- standard strain tensor).
\item
\ $\mathcal{M}=\mathcal{B}=M_4$ --- pseudoeuclidian Minkowski space, $\Theta$
--- Minkowski metrics. Such $d$-structure realizes relativistic generalization
of 3D elasticity theory \cite{kok4,ant}. Here
$\mathrm{MOT}_{M_4}(M_4)=\mathrm{P}^{\uparrow}_{+}(1,3)$
--- Poincare group with homogeneous subgroup of proper or\-tho\-chro\-nal Lorentz transformations.
\item
$\mathcal{M}=\mathcal{B}=V_4$ --- arbitrary 4D Riemannian manifold with
Riemannian metric $g=\Theta.$ Here we have generally covariant 4D elasticity
theory, which takes into account gravitational field. The $\mathrm
{MOT}_{V_4}(V_4)$ is isometry group of $V_4.$ Note, that while in two previous
cases $\mathcal{M}$ --- completely deformationally homogeneous, $V_4$ is, if
and only if it is homogeneous (in common sense).
\end{enumerate}

Lets note also the following less direct and obvious realization of free
$d$-structures.
\begin{enumerate}
\setcounter{enumi}{3}
\item
$\mathcal{M}$ --- $2n$-dimensional phase space of some dynamical system with
canonical symplectic $2-$form $\omega=\Theta$ \cite{arnold}.  If $d$-body
$\mathcal{B}\subseteq\mathcal{M}$
--- some closed subset of initial data, then
$\mathrm{MOT}_{\mathcal{M}}(\mathcal{B})$
--- Hamiltonian phase flow, going thro\-ugh  $\mathcal{B},$ which is (locally) generated by
some Hamiltonian function $h$ (while
$\mathrm{DEF}_{\mathcal{M}}(\mathcal{B})_0$
--- group of arbitrary proper diffeomorphisms of $\mathcal{M},$ generally speaking,
changing form $\omega$ (see Sec.\ref{maxwell})).  Dynamical systems with
constraints $\{f_a(p,q)=0\}$ are described by $d$-bodies
--- submanifolds of $\mathcal{M}$ and in this case
\[\mathrm{MOT}_{\mathcal{B}}(\mathcal{B})=\mathrm{MOT}_\mathcal{\mathcal{M}}(\mathcal{B})\cap
{\rm Sl}(\mathcal{B})\] and is defined by equation of motion for constraints
\[\{f_a,h\}=\sum\limits_a c_af_a,\] where $\{\ ,\ \}$ --- Poisons's brackets. Quite
different physical interpreting  of  symplectic $d$-structures,
$\mathrm{MOT}_{\mathcal{M}}(\mathcal{B})$ and
$\mathrm{DEF}_{\mathcal{M}}(\mathcal{B})_0$ we'll consider within dynamical
case in Sec.\ref{maxwell}.
\item
Lets $P(B,G)$ --- bundle space with base $B\stackrel{\pi}{\leftarrow}P$,
canonical projection $\pi,$ and structural group $\mathcal{G}$ \cite{kobayasi}.
Connection on $P$ can be defined by the 1-form
$\Theta\in\Omega(TP,\mathfrak{g}),$ which maps vector fields on $TP$ into Lie
algebra $\mathfrak{g}$ of $\mathcal{G}.$ So, $\varmt{P}{P}$ will consist of all
such deformations, which leave $\Theta$ invariant.  Note, that $\varmt{P}{P}$
is always nonempty, since under vertical diffeomorphisms $P\times G\to P$
$\Theta$ is invariant by its definition. If there are additional deformations,
which conserve $\Theta,$ it is said, that $\Theta$ is {\it invariant
connection.} Summary of some results on invariant connections for the case
$\varmt{P}{P}$ can be found in \cite{kobayasi}. Our approach requires more
general consideration in the case $\varmt{P}{\mathcal{B}}$ for $\mathcal{B}\neq
P.$

Another form, appearing in bundle space is 2-form of curvature: $\Omega\equiv
d\Theta+\Theta\wedge\Theta.$ So, formally, we could view on $\langle P, P,
\Diff{P},\{\Theta,\Omega\}\rangle$ as bimetric structure, but easily to show,
that $\Omega\preceq\Theta.$
\item
\ Let $\mathcal{M}$ --- space of all thermodynamical parameters, $\Theta\equiv\
Q$ --- heat form, $\mathcal{S}\equiv\mathcal{B}\subset\mathcal{M}$
--- some thermodynamical system, described by some set of equations of state $\{\varphi_\alpha=0\}.$
In other words, $\mathcal{B}$ is some admissible submanifold  in $\mathcal{M},$
which the system can evolve along.
 Then,
$\vardf{\mathcal{M}}{\mathcal{M}}_0$ describes arbitrary continuous evolutions
of thermodynamical parameters --- some given thermodynamical processes,
$\vardf{\mathcal{M}}{\mathcal{B}}_0,$ generally speaking,  describes continuous
changings of properties of the thermodynamical system (parameters of states
equation); $\mt{\mathcal{M}}$ --- continuous processes, conserving heat power,
$\mt{\mathcal{M}}\cap{\rm Sl}(\mathcal{B})$
--- continuous variations of state of the system, which conserves heat power,
$\mt{\mathcal{B}},$  --- continuous variations of properties, conserving heat
power. Continuous quasistatic changings of state of the system are exactly
proper slidings ${\rm Sl}(\mathcal{B}):\ \mathcal{B}\to\mathcal{B}.$
\end{enumerate}

\section{Dynamical deformational structures}\label{dyn}

\subsection{Definitions}\label{defdin}

We have developed the theory of free deformational structures, containing some
kinematical aspects  of the deformational  approach. To consider dynamics it is
necessary to supply a free structure $\mathfrak{D}$ with some {\it variational
principle $\mathfrak{A}.$} We define $\mathfrak{A}$ as the triad
$\langle\mathcal{F},\mu,\Gamma\rangle,$ where $\mathcal{F}:\
\tm{\Omega^{\otimes p}(\mathcal{B})}{\mathcal{B}}\to\mathbb{R}$ --- {\it scalar
energy density}, $\mu$ --- some {\it volume measure on $\mathcal{B}$}, $\Gamma$
--- {\it boundary conditions collection}. We'll call $\langle\mathfrak{D},\mathfrak{A}\rangle$
{\it dynamical $d$-structure} or simply $d$-{\it structure.} Lets discuss every
of $\mathfrak{A}$ components separately.

\begin{enumerate}
\item
Within standard continuum media physics dependence of $\mathcal{F}$ on
deformable bodies properties, on deformations  and on external conditions is
defined by, a so called, {\it material} or {\it definitional
relation}\footnote{In \cite{trus} material relations were defined as expressing
instant stress tensor $\sigma^t$ through {\it prehistory of system} $F_{t'},$
$t'\le t$. There is no importance how the relation is defined: by
$\sigma^t[F_{t'}]$ or by local energy $\mathcal{F}[F_{t'}]$ in view of the
relation $\sigma^t=\delta\mathcal{F}/\delta\Delta^t$. The second will be more
convenient for us.} and specified either by experiments or by some theoretical
considerations, such as {\it reference frame independence} (see \cite{trus}).
In present paper we restrict ourself by those $d$-bodies, whose definitional
relation 1) does not depend on "past prehistory" of deformations and 2) admits
the separation:
\[ \mathcal{F}=\mathcal{F}_0+\tilde U,
\]
where $\mathcal{F}_0$ -- {\it elastic part} -- depends only on deformation
measure, $\tilde U:\ \mathcal{B}\to\mathbb{R}$ -- {\it external potential part}
-- does not depend on  deformation form\footnote{In fact, $\tilde
U=\iota_\ast(U),$ where $U:\ \mathcal{M}\to\mathbb{R},\ \iota\in\mathcal{E},$
--- potential energy of $\mathcal{B}$ in external fields on $\mathcal{M},$ which can possess by
their own deformational dynamics.}. In analogy with similar common bodies,
satisfying the condition 1,  we'll call  such $d$-bodies {\it elastic} and
satisfying condition 2 --- {\it simple} and corresponding $d$-structures
--- {\it elastic} and {\it simple} respectively. We'll say, that $d$-structure is {\it closed,}
if  $\tilde\mathcal{U}\equiv{\rm const}_{\mathcal{B}\times I},$ and is {\it
open}, if $\tilde U\not\equiv{\rm const}_{\mathcal{B}\times I}.$ Everywhere
below we'll consider
$\mathcal{F}_0=\mathcal{F}_0(\Delta_\mathcal{B},\dot\Delta_\mathcal{B}),$ that
also restricts a wide class of {\it minimal $d$-structures} within more general
{\it high} ones, where $\mathcal{F}_0$ can depend on high derivatives of
$\Delta_{\mathcal{B}}.$
\item
Although the mapping $(d\iota)^{\ast}$ induces form
$(d\iota)^\ast\Theta\equiv\Theta_\mathcal{B}$ on $\mathcal{B},$  there is a
problem to build local volume form $d\mu$ and scalar $\mathcal{F}_0(\Delta),$
besides the case\footnote{Even under $p=2$ one should check, that ${\rm
det}\,\Theta_{\mathcal{B}}\not\equiv0$ (see Appendix \ref{inverse}).} $p=2.$ We
have the following two alternatives: 1) try to define somehow form $d\mu$ and
scalar $\mathcal{F}_0(\Delta)$ in terms of $\Theta_\mathcal{B}$ in case of
general forms $\Theta_{\mathcal{B}}\in\Omega^{\otimes p}(\mathcal{B})$; 2) to
define $d\mu$ and (or) $\mathcal{F}_0(\Delta)$ on $\mathcal{B}$ independently
on $d$-metrics $\Theta_{\mathcal{B}}.$ $d$-structures, realizing the first
alternative will be called {\it internal}, second --- {\it external}. Present
paper will be mainly concerned with (more economical) internal $d$-structures
(see Sec.\ref{int} and Appendix \ref{vol}). For future purposes we'll call the
metrics, which define scalar products and (or) volume form {\it $g$-metrics,}
in difference with $d$-metrics, defining deformation measure.
\item
Boundary conditions we'll discuss and specify after derivation of
Euler-Lagrange equations in Sec.\ref{boundary}.
\end{enumerate}

\subsection{Static and evolutional cases}\label{cases}

In applications of the approach to different physical systems we'll be faced
with the two types of dynamical deformational structures --- {\it static} and
{\it evolutional}. To differ them we introduce special index $\epsilon,$ which
takes value "1" in case of evolutional  structures and "2" --- in case of
static ones. Variational functional $\mathfrak{F}$ can be written then as the
following universal expression: \begin{equation}\label{vargen}
\mathfrak{F}[\Upsilon_{\epsilon}]=\int\limits_{\mathcal{C}_{\epsilon}}(\mathcal{F}_0
(\Delta_{\epsilon})+\tilde\mathcal{U})\,d\mu_{\epsilon},
\end{equation}
where in notations of Sec.\ref{defree}, \ref{hom}, \ref{defdin}
\[
\Upsilon_1=F_t(\mathcal{B}),\ \mathcal{C}_1=\tm{\mathcal{B}}{I},\
\Delta_1=\{\Delta^t_\mathcal{B},\dot\Delta^t_\mathcal{B}\},\ \
d\mu_1=(d\pi_2)^\ast e(t)dt\wedge(d\pi_1)^\ast d\mu,\]
\[\Upsilon_2=\iota'(\mathcal{B}),\ \mathcal{C}_2=\mathcal{B},\
\Delta_2=\Delta_\mathcal{B},\ d\mu_2=d\mu.
\]
 Here $e(t)$
--- some "metric" on $I,$ $\pi_1,\pi_2$ --- projections of
$\tm{\mathcal{B}}{I}$ on the first and second multipliers respectively. In
other words, in case $\epsilon=1$ we find minimum of $\mathfrak{F}[F_t]$ and
vary evolution $F_t,$ while "ends points" $\{F_{\partial I}\}$ hold fixed. In
case $\epsilon=2$  we find minimum of $\mathfrak{F}[\iota'],$ varying final
embedding $\iota'$, while initial embedding $\iota$ hold fixed.

\subsection{Internal $d$-structures\\ and $g$-metrics}\label{int}

The fact of existence of scalar density $\mathcal{F}_0(\Delta)$ and variational
functional (\ref{vargen}) put some restrictions on possible kinds of
$d$-metrics $\Theta_\mathcal{B}.$ Within internal $d$-structures this metric
merely should:

1) admit the isomorphism $\Omega^{\otimes p}(\mathcal{B})\to V^{\otimes
p}(\mathcal{B}),$ where $V^{\otimes p}(\mathcal{B})\subset T(0,p)(\mathcal{B})$
--- subspace of contravariant tensor fields  of valency $p$ ($p$-vectors).
With the help of the isomorphism we are
 able   to build  from $\Delta$
scalars of type $(\Delta,\Delta)\equiv\langle\Delta,\tilde\Delta\rangle,$ where
$\tilde\Delta\in V^{\otimes p};$

2) provide possibility for constructing of invariant volume form $d\mu=\varpi\,
dx^1\wedge\dots\wedge dx^d,$  where $\varpi=\varpi(\Theta)$
--- scalar density of weight $-1$ with respect to coordinate diffeomorphisms on $\mathcal{B}.$

In Appendices (\ref{inverse})-(\ref{induce}) we generalize standard square
matrix calculus on arbitrary  form of even degree $p=2k.$ The results are
following:

\begin{enumerate}
\item
{\it A form $\Theta_\mathcal{B}$  of even degree $2k$ admits point-wise
isomorphism  $\Omega^{\otimes 2k}\to V^{\otimes 2k}$ and has inverse
$2k-$vector $\Theta^{-1}_{\mathcal{B}}$:
$\Theta_{\mathcal{B}}^{-1}\cdot\Theta_{\mathcal{B}}=\Theta_{\mathcal{B}}\cdot\Theta_{\mathcal{B}}^{-1}=E$,
if in some coordinate system its matrix  is point-wise preimage of  the
isomorphism $\chi_\ast$ of any nondegenerate section of trivial bundle
$\mathcal{B}\times\mathrm{M}_{d^k\times d^k}(\mathbb{R})$:}
\[
\|\Theta_{\mathcal{B}}\|=\chi_\ast^{-1}(M(\mathcal{B})),\
M(b)\in\mathrm{M}_{d^k\times d^k}(\mathbb{R}),\ \ {\rm  det}\, M(b)\neq0,\
\forall\ b\in\mathcal{B},
\] where  $\chi_\ast$ and all another notations are introduced in Appendix
\ref{inverse}.
\item
{\it For any natural  $d$ and  $k,$ related by the equation
\[ 3^k-2kd^{k-1}=4m+1,\ m\in\mathbb{Z}\]
 there exists volume form on  $\mathcal{B}$ of the kind:}
\[
\left|\overline{\rm
det}\,\Theta_\mathcal{B}\right|^{1/2kd^{k-1}}dx^1\wedge\dots\wedge dx^d
\equiv\left|{\rm
det}\,\chi_\ast(\Theta_{\mathcal{B}})\right|^{1/2kd^{k-1}}dx^1\wedge\dots\wedge
dx^d,
\]
{\it where $\Theta_\mathcal{B}$ --- any form of degree $2k$, satisfying the
existence of $\Theta_{\mathcal{B}}^{-1}$ condition.}
\item
{\it The form $\Theta_\mathcal{B}$ as image $(d\iota)^\ast\Theta_\mathcal{M}$
is nondegenerate, if and only if}
\begin{equation}\label{cond1bis}
(L_{\Theta(\mathcal{M})}\cup R_{\Theta(\mathcal{M})})\cap
(T\mathcal{S})^{\otimes k}=\varnothing.
\end{equation}
or in words, {\it when left and right kernels of the form $\Theta_\mathcal{M}$
has null intersections with the space of all $k-$vector $V^{\otimes
k}(\mathcal{S})$ on a whole $\mathcal{S}$} (see Appendix \ref{induce}).
\end{enumerate}

Everywhere below we assume, that $d$-metric has even valency and satisfies all
conditions 1,2,3.

\section{$d$-objects  dynamical (equilibrium) equations}\label{eqsc}

Now we are going to derive general dynamical equation  of $d$-objects. Lets
introduce some useful indexless matrix notations, adopted both for static and
for evolutional problems.

\subsection{Description of embeddings and deformation measure}\label{descr}

  We'll describe some history $F_t(\mathcal{B})$  by the set of
functions\footnote{They are often called in literature {\it embedding
variables.}} $\{x^A(\xi,t)\}_{A=1,\dots,n},$ where
\[\{x^A\}_{A=1,\dots,n},\
\{\xi^{\alpha},t\}_{\alpha=1,\dots,d}\] --- coordinates on $\mathcal{M}$ and
$\mathcal{B}\times I$ respectively. This multicomponent notation we'll short as
usually  to $x=x(\xi,t)\equiv x_t(\xi).$ Corresponding matrix $Dx_t$ for
$(dF_t)^\ast$  has the components \[
\left(Dx_t\right)^{A}_{\alpha}=\frac{\partial x_t^{A}}{\partial\xi^\alpha}.\]
Codifferential $(dF_t)^\ast$  defines induced linear mapping:
\begin{equation}\label{inducec}
L_t\equiv(Dx_t)^{\otimes p}
\end{equation}
of $d$-metric --- representation of $(dF_t)^\ast$ in
$(T^\ast\mathcal{S})^{\otimes p},$ such that:
\[
\Theta^t_{\mathcal{B}}=(dF_t)^\ast\Theta\equiv L_t\Theta_{\mathcal{M}}.
\]

For  measure of deformation we have in evolutional case:
\begin{equation}\label{not}
\Delta^t_{\mathcal{B}}=(dF_{t})^{\ast}\Theta- (dF_{0})^{\ast}\Theta\equiv
L_t\Theta(x_t)-L_0\Theta(x_0)\equiv\Theta^t_{\mathcal{B}}-\Theta^0_{\mathcal{B}};\
\
\dot\Delta_{\mathcal{B}}\equiv\dot\Theta^t_{\mathcal{B}}=\frac{d\Theta_{\mathcal{B}}^t}{dt},
\end{equation}
where $\Theta^0_{\mathcal{B}}\equiv L_0\Theta$ --- {\it background (initial)}
metric.

For static problem we formally put: \[\Delta_{\mathcal{B}}=
(dF_{1})^{\ast}\Theta- (dF_{0})^{\ast}\Theta\equiv
L_1\Theta(y)-\Theta^0_{\mathcal{B}},\] where $y\equiv x_1(\xi).$

\subsection{Equations of motion (evolutional case)}\label{eq}

Lets begin from a more general evolutional case. Accordingly to Sec.\ref{cases}
(case $\epsilon=1$) full action has the following kind:
\begin{equation}\label{full}
\mathfrak{F}[x_t(\xi)]=\int\limits_{\tm{\mathcal{B}}{I}}
(\mathcal{F}_0+\mathcal{U})\,\upsilon dt\wedge d\mu_\xi,
\end{equation}
where
$\mathcal{F}_0=\mathcal{F}_0(\Delta^t_{\mathcal{B}},\dot\Delta^t_{\mathcal{B}})$
--- internal elastic part of energy of (generally speaking,
nonhomogeneous\footnote{Such nonhomegeneous $d$-body possess different elastic
properties at different points. We should denote it by using apparent
dependency of $\mathcal{F}$ on $\xi,$ but for the brevity don't do it.})
$d$-body $\mathcal{B}$, $\mathcal{U}(x_t(\xi))\equiv\tilde \mathcal{U}(\xi,t)$
--- external potential part of energy, $\upsilon\equiv
e(t)\cdot\varpi(\Theta^t_\mathcal{B}),$ $e\,dt$
--- metric on $I,$ $\varpi\, d\mu_\xi$
--- volume form on $\mathcal{B},$ induced by $\Theta^t_\mathcal{B},$
$d\mu_\xi\equiv d\xi^1\wedge\dots\wedge
 d\xi^d.$

First variation of (\ref{full}) over $x_t(\xi)$ takes the form:
\[
\delta\mathfrak{F}=\int\limits_{\tm{\mathcal{B}}{I}}
\left\{(\delta\mathcal{F}+\delta\mathcal{U})\upsilon
+(\mathcal{F}+\mathcal{U})\delta\upsilon\right\}dt\wedge d\mu_\xi. \]
Everywhere below in our derivation we'll omit  $\mathcal{B}$ and $t$ at the
bottom and top of $\Delta$ and of other values. Using the relations and
definitions:\begin{equation}\label{str} \delta \mathcal{F}=
\langle\mathcal{F}_{|\Delta},\delta\Delta\rangle+\langle\mathcal{F}_{|\dot\Delta},\delta\dot\Delta\rangle
\equiv \langle\sigma,\delta\Delta\rangle+\langle\pi,\delta\dot\Delta\rangle;\ \
\delta\mathcal{U}=\langle\mathcal{U}_{|x},\delta x\rangle;\ \
\delta\upsilon=\langle\upsilon_{|\Delta},\delta\Delta\rangle,
\end{equation}
 where we have
introduced  {\it stress tensor} $\sigma$ and {\it surface momentum density
tensor} $\pi,$ have used  $\langle\ ,\ \rangle$ for coordinateless
representation of summation as "linear functional" over variations in
corresponding spaces and have taken into account, that by (\ref{not})
$\delta\Delta^t_{\mathcal{B}}=\delta\Theta^t_{\mathcal{B}}.$

After integrating by parts over $t$ we have:
\[
\delta \mathfrak{F}=\int\limits_{\tm{\mathcal{B}}{I}} \{ \langle
\sigma\upsilon-\frac{d}{dt}(\pi\upsilon)+
(\mathcal{F}+\mathcal{U})\upsilon_{|\Delta}\,,\,\delta\Delta\rangle
+\langle\mathcal{U}_{|x}\upsilon,\delta x\rangle \}\,dt\wedge d\mu_\xi+
\int\limits_{\tm{\mathcal{B}}{\partial
I}}\langle\pi,\delta\Delta\rangle\upsilon\, d\mu_\xi.
\]
The first triangle bracket within volume term  can be transformed by the
following way:
\[
\sigma\upsilon-\frac{d}{dt}(\pi\upsilon)+(\mathcal{F}+\mathcal{U})\upsilon_{|\Delta}=\sigma\upsilon-\dot\pi\upsilon-\pi\langle\upsilon_{|\Delta},\dot\Delta\rangle+
(\mathcal{F}+\mathcal{U})\upsilon_{|\Delta} =(\sigma-\dot\pi)\upsilon
-\hat\mathcal{T}\upsilon_{|\Delta}\equiv\upsilon\Sigma,
\]
where we have introduced
\[\hat\mathcal{T}\equiv\pi\otimes\dot\Delta-(\mathcal{F}+\mathcal{U})\hat I\] ---
{\it deformational  energy-momentum affinnor,} $\hat I$ --- identical linear
operator in $T(0,p)$: $\hat I\,v_{|\Delta}=v_{|\Delta},$
\[\Sigma\equiv(\sigma-\dot\pi-\hat\mathcal{T}\,(\ln\varpi)_{|\Delta})\] --- {\it
generalized} stress tensor.

Simple calculation with using (\ref{inducec}) and (\ref{not}) gives:
\begin{equation}\label{varinduce}
\delta\Delta=\delta(L\Theta)=\delta L\Theta+L\delta\Theta
\equiv\langle\check{\Theta},\delta
Dx\rangle+\langle(\Theta_{|x})_\mathcal{B},\delta x\rangle,
\end{equation}
where
\[\check{\Theta}_{\!\!\underbrace{\scriptstyle\alpha_1\dots\alpha_p}_{\scriptscriptstyle p-1\
\mbox{\tiny\rm indexes}}\!\! A}\equiv\left( \frac{\partial L}{\partial(Dx)}
\right)_{\!\!\underbrace{\scriptstyle\alpha_1\dots\alpha_p}_{\scriptscriptstyle
p-1\ \mbox{\tiny\rm indexes}}\!\! A}\equiv\sum\limits_{s=1}^{p}
{\Theta_\mathcal{B}}_{\alpha_1\dots\alpha_{s-1}A\alpha_{s+1}\dots\alpha_p},\ \
(\Theta_{|x})_\mathcal{B}\equiv L\Theta_{|x}.\] Substituting all into
$\delta\mathfrak{F}$ and integrating by parts over $\xi,$ we have:
\[
\delta\mathfrak{F}= \int\limits_{\tm{\mathcal{B}}{I}} \{ \langle
 -\partial_\xi(
\upsilon\widetilde{\Sigma})+ \langle\upsilon\Sigma,(\Theta_{|x})_{\mathcal{B}}
\rangle+\upsilon\mathcal{U}_{|x},\delta x\rangle \} \,dt\wedge d\mu\]
\begin{equation}\label{varfin}
+ \int\limits_{\tm{\mathcal{B}}{\partial
I}}\langle\pi,\delta\Delta\rangle\upsilon\,
d\mu_\xi+\int\limits_{\tm{\partial\mathcal{B}}{I}}\langle\widetilde{\Sigma},\delta
x \rangle\,\upsilon dt\wedge d\mu'_\xi,
\end{equation}
where we use notation $\widetilde\Sigma=\langle \Sigma,\check\Theta\rangle$ and
in last boundary integral $d\mu'_\xi$ symbolizes elements of the sets
$\{d\xi^{\alpha_1}\wedge\dots\wedge d\xi^{\alpha_{d-1}}\}$ of $d-1$ coordinate
boundary hypersurface volume form.

Extremality condition $\delta\mathfrak{F}=0$ gives the following equations of
motion:
\begin{equation}\label{eqeq}
{\rm div}\,\Sigma+f_{\Theta}+f_{\rm ext}=0,
\end{equation}
where
\[
({\rm div}\, \Sigma)_A\equiv\frac{1}{\upsilon}\partial_{\alpha_s}
\left(\upsilon\Sigma^{\alpha_1\dots\alpha_s\dots\alpha_p}\check\Theta_{\alpha_1\dots\alpha_pA}\right)
\equiv
\frac{1}{\upsilon}\partial_{\alpha}\left(\upsilon\widetilde{\Sigma}^{\alpha}_A\right)\equiv
\frac{1}{\varpi}\partial_{\alpha}\left(\varpi\widetilde{\Sigma}^{\alpha}_A\right)\]
--- operator of divergence,
\[
f_{\Theta}\equiv-\left\langle\Sigma,(\Theta_{|x})_{\mathcal{B}}\right\rangle
\]
--- {\it $\Theta$-gravity force density,} induced by nonhomogeneity of $\Theta$
in $\mathcal{M}$ (it vanishes, when $\Theta$ --- constant form),
\[
f_{\rm ext}\equiv-\mathcal{U}_{|x}
\]
{\it --- external force density}, induced by external fields (it vanishes in
case of closed $d$-structures).

\subsection{Boundary conditions}\label{boundary}

Under derivation of dynamical equations we have obtained boundary conditions
(\ref{varfin}) of the following general kind:
\begin{equation}\label{genb}
\int\limits_{\Gamma_a}\left\langle X_a,\delta x\right\rangle\, d\mu_a,\ \ \
a=1,2,3,
\end{equation}
having sense of vanishing of "average work" on variations $\delta x$ at
boundary. Here boundary\footnote{The term with $\Gamma_3$ arises after
integrating by parts of term with $\Gamma_2$ with using (\ref{varinduce}).}
\[\Gamma=\partial(\mathcal{B}\times I)=(\partial\mathcal{B}\times
I)\cup(\mathcal{B}\times\partial I)\equiv\Gamma_1\cup\Gamma_2;\ \
\Gamma_3=\Gamma_1\cap\Gamma_2=\partial\mathcal{B}\times\partial{I};\]
\[X_1=\widetilde\Sigma;\ \ X_2=-{\rm
div}\,\pi+\langle\pi,(\Theta_{|x})_{\mathcal{B}}\rangle;\ \
X_3=\widetilde\pi\equiv\langle\pi,\check\Theta\rangle;\]
\[d\mu_1=\upsilon\,dt\wedge d\mu'_{\xi};\ \
 d\mu_2=\upsilon d\mu_\xi;\ \ d\mu_3=\upsilon d\mu'_\xi.
\]

Here we consider only the most known and widely used boundary conditions
(generalizing ones in standard elasticity theory):
\begin{enumerate}
\item {\it Pinned boundaries $(P)$.} In this case $\delta x|_{\Gamma}=0$ and
all equations (\ref{genb}) are satisfied identically.
\item {\it Free boundaries
$(F)$.} In this case variations $\delta x$ are arbitrary on $\Gamma$ and
boundary conditions takes the form:
\[
(F):\ \ X_a|_{\Gamma_a}=0\ \ \ a=1,2,3.
\]
One only should check consistency of this independent equations on $\Gamma_3.$
\item {\it Sliding boundaries $(S)$.} Let $\delta
x_\tau\equiv\delta F_\tau(\Gamma)$ --- variational homotopy of $\Gamma.$ We'll
relate  $\delta x_\tau$ to a class of variations of {\it sliding type $\delta
x_{\parallel}$,} if $d\delta F_\tau(\widetilde{d/d\tau})\in T\Gamma.$ Then
sliding boundary conditions takes the form:
\[
(S):\ \ \  \left\langle X_a,\delta x_{\parallel}\right\rangle=0,\ \ a=1,2,3.
\]
If $\{\eta^\alpha\}$ --- coordinates on  $\Gamma,$ then its image
$\iota(\Gamma)$ in $\mathcal{M}$  can be described by the set of functions
$\{x(\eta)\}.$ The set $\{\partial_\eta x\}\subset T\mathcal{M}$ forms
collection of basis vector fields on $\iota(\Gamma).$ Then coordinate form of
sliding boundary conditions will be
\[
(S):\ \ \ \left\langle X_a,\partial_\eta x\right\rangle=0,\ \  a=1,2,3. \]  One
should only check it consistency on $\Gamma_3.$
\item {\it Boundaries with given  variations $(R)$.} If $\delta x|_{\Gamma_a}\equiv\varphi_a(\eta)$ ---
some fixed functions on $\Gamma,$ such that
$\varphi_3=\varphi_2|_{\Gamma_3}=\varphi_1|_{\Gamma_3}$ then we come back to
general conditions (\ref{genb}) and get:
\[
(R):\ \ \int\limits_{\Gamma_a}\left\langle X_a,\varphi_a\right\rangle\,
d\mu_a=0,\ \ a=1,2,3.\]
\end{enumerate}

\subsection{Static case}\label{steq}

In the static case accordingly to the Sec.\ref{cases} (case $\epsilon=2$) we
start from the action:
\[
\mathfrak{F}[y(\xi)]=\int\limits_{\mathcal{B}}(\mathcal{F}_0+\mathcal{U})\,
\varpi d\mu_\xi.
\]
Then we should carry out similar to the evolutional case manipulations, that
lead to the  particular case of  evolutional equations (\ref{eqeq}) and
boundary conditions (\ref{genb}), taken under
\[
\pi=0,\ \ I=\{0,1\},\
\
\partial I=\varnothing,\ \ \upsilon\to\varpi.
\]
So, in static case we obtain (\ref{eqeq}) as equilibrium equations, with
\[
\hat\mathcal{T}=-(\mathcal{F}_0+\mathcal{U})\hat I;\ \
\Sigma=\sigma-\hat\mathcal{T}(\ln(\varpi))_{|\Delta}
\]
and with the only boundary condition:
\[
\int\limits_{\partial\mathcal{B}}\langle\widetilde\Sigma,\delta x\rangle\,
\varpi d\mu'_\xi=0.
\]

\subsection{Perturbative elasticity theory}\label{pert}

Since deformational energy density  $\mathcal{F}_0(\Delta)$ is
scalar\footnote{Here we don't differ $\Delta$ and $\dot\Delta$ since they
possess similar algebraic structure. So, our consideration in this paragraph
touches pure cases $\mathcal{F}_0(\Delta_\mathcal{B})$ or
$\mathcal{F}_0(\dot\Delta)_\mathcal{B},$ but general mixed case
$\mathcal{F}_0(\Delta,\dot\Delta)$ can be considered by the similar manner}, it
can depend on $\Delta$ only through the following combinations:
\begin{equation}\label{comb}
\Delta^{(i)}\equiv {\rm Tr}\,(\bar\Delta^i),
\end{equation}
where $\bar\Delta\equiv\Delta\cdot(\Theta^0)^{-1}\in T(p/2,p/2)$ and matrix
degree, multiplication and  trace operation are  understood in the sense of the
corresponding operations of its $\chi_{\ast}$-images (see Appendix
\ref{inverse}) in some coordinate system. Since ${\rm dim}\, \mathcal{B}=d,$
then there exists no more then $d$ fun\-cti\-o\-nal\-ly-independent scalars
$\Delta^{(i)},$ which can be ordered by increasing $i.$ Let
$\{\Delta^{(i_{l})}\}_{1\le i_l\le I;\ l=1,\dots,s\le d}$ will be collection of
the first $s$ such independent scalars.

To compare equations of the Sec.\ref{eq},\ref{steq} with well known equations
of standard field theory it is necessary to go to decomposition of the energy
$\mathcal{F}_0$ over power of $\Delta.$ We'll  see, that the most part of
modern field-theoretical models can be described by the first members of the
decomposition --- the so called $\Delta^1$ and $\Delta^2$ -structures (see
below).
%Lets consider functional  $\mathcal{}[\zeta]$ from the viewpoint of
%perturbative field theory for the deformation $\zeta$ which are close to the
%rigid motions.
So we need investigate the structure of the following formal row:
\begin{equation}\label{row}
\mathcal{F}_0(\Delta)=
\sum\limits_{i=1}^{\infty}\frac{1}{i!}\left.\frac{\partial^i\mathcal{F}_0}{\partial\Delta^i}\right|_{\Delta=0}\Delta^i.
\end{equation}
The symbolic Macloren row (\ref{row}) with using  notations  (\ref{comb}) can
be rewritten as follows:
\begin{equation}\label{decomp}
\mathcal{F}_0= \sum\limits_{i=0}^{\infty}\sum_{(\vec k,\vec
i)=i}\mu^{i}_{k_1\dots k_i}
(\Delta^{(i_1)})^{k_1}(\Delta^{(i_2)})^{k_2}\cdots(\Delta^{(i_s)})^{k_s},
\end{equation}
where in the second sum there is summation over all vectors $\vec
k=(k_1,\dots,k_s)$ of $s-$dimensional integer-valued lattice, whose nonnegative
coordinates satisfy the equation of atomic hyperplane $(\vec k,\vec i)=i.$
Parenthesis denote Euclidean scalar product in $\mathbb{E}^s,$ the vector $\vec
i=(i_1,i_2,\dots,i_s).$ Scalar coefficients $\{\mu^{i}_{k_1\dots k_i}\}$
characterize  "elastic properties" of the $d$-body. Similarly to standard
elasticity theory we'll call it  {\it generalized Lame coefficients.}

We'll call deformational structure $\mathfrak{D}$  with energy density
$\mathcal{F}_0$ as exact  finite sum of powers of $\Delta$ with highest term of
order $i$ in (\ref{decomp}) {\it $\Delta^i$-structure.}

Lets consider in more details  $\Delta^3$-structure, assuming that the scalars
$\Delta^{(1)},\Delta^{(2)},\Delta^{(3)}$ are nonzero and independent.

 1) $i=0.$ There is one Lame coefficient $\mu^0\equiv\mathcal{F}_0(0),$
which represent background (null) energy and practically always can be
annihilated by constant shift of $\mathcal{F}_0.$

 2) $i=1.$ There is one Lame coefficient $\mu^1_1.$
The corresponding term of finite sum is:
\begin{equation}\label{term}
\mu^1_1\Delta^{(1)}.
\end{equation}
Within the standard elasticity theory the term is responsible  for energy of
strongly tensed bars and plates (strings and membranes) and  heat expanding of
isotropic bodies \cite{land}.

 3) $i=2.$ There is two Lame coefficient $\mu^2_{01}$ and
 $\mu^2_{20}$
and two terms in $\mathcal{F}_0$ respectively:
\begin{equation}\label{Hook}
\mu^2_{01}\Delta^{(2)}+\mu^2_{20}(\Delta^{(1)})^2.
\end{equation}
The expression is well known {\it Hooks law} of linear elasticity theory (where
$\mu^2_{01}=\mu$ --- shift modulus, $\mu^2_{20}=\lambda/2$ --- second
independent Lame coefficient).

4) $i=3.$ Three nonzero Lame coefficient $\mu^{3}_{001},$ $ \mu^3_{110},$ $
\mu^3_{300}$ gives the following terms in $\mathcal{F}_0$:
\[\mu^3_{001}\Delta^{(3)}+\mu^3_{110}\Delta^{(1)}\Delta^{(2)}+\mu^3_{300}(\Delta^{1})^3.\]
This part  describes nonlinear corrections to the linear models within
elasticity and field theory. In present paper  we'll not touch it.

So, we have the following  general kind  of $\mathcal{F}_0$ within
$\Delta^3$-structure:
\begin{equation}\label{energy3}
\mathcal{F}_0(\Delta^{(1)},\Delta^{(2)},\Delta^{(3)})=\mu^0+\mu^1_1\Delta^{(1)}
+\mu^2_{01}\Delta^{(2)}+\mu^2_{20}(\Delta^{(1)})^2+
\mu^3_{001}\Delta^{(3)}+\mu^3_{110}\Delta^{(1)}\Delta^{(2)}+\mu^3_{300}(\Delta^{1})^3.
\end{equation}
Lets calculate  stress tensor\footnote{Surface momentum density can be obtained
by changing $\Delta\to\dot\Delta$.} $\sigma$ for $\Delta^3$-structure. Using
its definition in (\ref{str}) and decomposition (\ref{energy3}), we have:
\[
\sigma\equiv\frac{\partial\mathcal{F}_0}{\partial\Delta}=
\sum\limits_{i=1}^3\frac{\partial\mathcal{F}_0}{\partial\Delta^{(i)}}
\frac{\partial\Delta^{(i)}}{\partial\Delta}
=(\mu^1_1+2\mu^2_{20}\Delta^{(1)}+\mu^3_{110}\Delta^{(2)}+3\mu^3_{300}
(\Delta^{(1)})^2)(\Theta^0)^{-1}\]
\begin{equation}\label{stress3}
+2(\mu^2_{01}+\mu^3_{110}\Delta^{(1)})(\Theta^0)^{-1}\cdot\Delta\cdot(\Theta^0)^{-1}
+3\mu^3_{001}
(\Theta^0)^{-1}\cdot\Delta\cdot(\Theta^0)^{-1}\cdot\Delta\cdot(\Theta^0)^{-1}.
\end{equation}
The expression (\ref{energy3}), (\ref{stress3}) generalize in many aspects well
known expression for elastic energy and stresses tensor within standard linear
elasticity theory \cite{land}.

\section{Examples of dynamical deformational structures}\label{exam}

Now we consider  some examples of classical $d$-stru\-ctu\-res, which can be
observed within the well known theories. We leave without attention those
examples, which concern either with standard elasticity theory --- starting
point of our generalizations, or with its development in $M_4$ or in $V_4,$
mentioned in Sec.\ref{example1}, since we are intending to devote them special
papers in future.

\subsection{Example 1: The theory of classical  $d-1-$brane}\label{example2}

Let $\mathcal{M}=M_{N+4}$ be pseudoeuclidian space with metric $\Theta.$ Lets
consider static $\Delta^1$-structure with arbitrary $d$-body $\mathcal{B}.$
There is unique scalar invariant: \[ \Delta^{(1)}={\rm
Tr}[(\Delta\cdot(\Theta^0)^{-1}]={\rm Tr}(Dy^{T}\cdot \Theta\cdot Dy
-\Theta^0)\cdot(\Theta^0)^{-1})
 =\Theta\otimes(\Theta^0)^{-1}(Dy,Dy)-d\equiv|Dy|^2-d,\] where
$\Theta\otimes(\Theta^0)^{-1}$ contracts $Dy$ as vectors in $TM_{N+4}$ and as
forms in $T^\ast\mathcal{\mathcal{B}}.$ Assuming $\mu^1_1=T/2$ in (\ref{term}),
we obtain the action of the following kind \footnote{Here we omit external
potential energy $\mathcal{U}.$ }:
\begin{equation}\label{string}
\mathfrak{F}=\frac{T}{2}\int\left(|Dy|^2-d\right)\sqrt{{\rm
det}\Theta^0}\,d\mu_{\xi}.
\end{equation}
This expression  coincides with well known Polyakov's action for classical
$d-1$-brane with special cosmological term \cite{green}. In a difference with
string and brane models the metric $\Theta^0$  is considered here as fixed
(background). In accordance with string and brane ideology variation of the
(\ref{string}) over $(\Theta^0)^{-1}$ leads to the constraint:
\[
\Theta(Dy,Dy)-\frac{1}{2}\left(|Dy|^2-d\right)\Theta^0=0. \] Its contraction
with $(\Theta^0)^{-1}$ leads to the relation:
\[|Dy|^2=\frac{d^2/2}{d/2-1},\] which under $d=2$ (string case) gives inconsistent
constraint $d=0.$ This arguments, typical for original string theory, are not
so catastrophic within deformational  approach, since true dynamical variables
are not components of metric $\Theta^0,$ but embedding  variables $x_0(\xi).$
If we  minimize $\mathfrak{F}$ with respect to both  final and initial position
of $d$-object, we obtain the following consistent system:
\[
{\rm div}\,Dy=0;\ \  {\rm div}\,(D x_0-2\Theta\otimes(\Theta^0)^{-1}(Dy,D x_0)D
y)=0,
\] where the first is obtained by $y-$variation (it is identical to the
string theory equation) and second --- by $x_0-$variation of action
(\ref{string}). We do not write here boundary conditions. Note also, that
cosmological term $-d\cdot T/2$ can be absorbed by suitable choice of
$\mathcal{F}_0(0).$

\subsection{Example 2: Classical
solids dynamics as
$\Delta^1_{\parallel}+\Delta^2_{\perp}$-structure}\label{example3}

Let $\mathcal{M}=M_4,$ $\Theta$ --- Minkowski metric, $\mathcal{B}\subset M_4$
--- thin 4D time-like bar, i.e. body, whose size along time-like
direction much more then in space-like. Within approach, been proposed in
\cite{kok4}, it performs the so called {\it absolute ("objective") history of
thing}, while its space-like sections,  observing from the point of view of
some reference frame, performs {\it relative ("subjective") history of the
thing.} Then, we have endowed the bar by some linear elastic properties,
described by Lame coefficients $\mu^2_{01}=\mu$ and $\lambda=2\mu^2_{20},$ (see
Hooks law (\ref{Hook})) and have generalized standard elasticity theory of
common bars in Euclidean 3D space on the 4D case. Analysis of the theory has
led to the following curious conclusions:

\begin{itemize}
\item
Classical mechanics can be formulated within 4D static deformational picture in
terms of straining of the thin strongly tensed bars (strings) without special
notion of mass (it has 4D force nature). In terms of deformational structures
such theory  should be related to the anisotropic
$\Delta^1_{\parallel}+\Delta^2_{\perp}$-structure, where symbols "$\parallel$"
and "$\perp$" differ time-like and space-like directions within our $d$-objects
--- 4D strings.   General formulation of such anisotropic $d-$stru\-ctu\-res is
obvious, but goes beyond the scope of the present paper.
\item
Only third Newton's law\footnote{As it has been cleared in \cite{trus}  third
Newton's law follows only from assumption of additivity of force function $f$
on bodies of mechanical Universe $\Omega$: $f(U_1\cup U_2)=f(U_1)+f(U_2),\
\forall \ U_1,U_2\subset\Omega.$} remains independent, while first and second
appear as its consequences. Curiously, that second Newton's law can be viewed
as 1D Laplace formula for strings, similar to 2D case for membrane.
\item
The approach reveals, that classical Newton laws are, in fact, result of some
extremely exact "tunings" in mechanical structure of Universe, which itself can
be imagined as {\it twisted and strongly tensed net}.
\item
In principle, there is possibility of violation of Newton dynamics in some
special situation: rapid rotations, large accelerations, beginning and end of
absolute history of some 3D body and others additionally to relativistic
effects.
\item
The approach reveals fundamental role of observer as not only "spectator" but
"participants" of  formation of physical laws even within classical mechanics
(see also (\cite{pav2})).
\end{itemize}

Similar ideas, revealing connections of elasticity and inertia has been
discussed in \cite{ant}. We hope that the deformational picture of classical
mechanics will provide useful means for its more deep understanding.

\subsection{Example 3: Einstein gravity as $\Delta^2$-structure}\label{example4}

Let as in Example 1  $\mathcal{M}=M_{N+4}$ --- pseudoeuclidian space with
metric $\Theta$ and $\mathcal{B}\subset M_{N+4}$
--- {\it thin 4D plate,} i.e. body, whose sizes along some four dimensions (one --- timelike and three
space-like)  are much more then in other ones. In the works
\cite{kok1,kok2,kok3} some generalization of standard elasticity theory for
common (2D in $\mathbb{E}^3$) plate has been applied for the 4D plate
equilibrium problem. We have endowed $\mathcal{B}$ with elastic constants
$\lambda=2\mu^2_{20}$ and $\mu=\mu^2_{01}$ and derived energy of bending
$\mathfrak{F}_{b}$ (\cite{kok1,kok2}) and stretching $\mathfrak{F}_s$
(\cite{kok3}) by integrating over extradimensions and 4D directions within
static $\Delta^2$-structure. We had found, that:
\begin{itemize}
\item  Theory of straining  of 4D plates in $M_{N+4}$ can  describe
space-time (this is plate itself!) and matter (this is special stresses of the
plate) dynamics in unified language. Namely, pure bending energy
$\mathfrak{F}_{b},$ calculated within $\Delta^2$-structure (up to a dimensional
constant) generalizes linearized Gilbert-Ein\-stein's action for gravity. On
the other hand, pure stretch energy $\mathfrak{F}_{s}$ plays role of action of
4D matter field, living on the plate.
\item
Within the deformational approach physical essence of Einstein equations
becomes very clear. They express vanishing of total 4D stres\-ses on the plate,
induced by bending and stretching. In other words, {\it Einstein equations
says, that true dynamic of space-time is realized as locally non\-stres\-sed
states of space-time.}
\item
This strange (from the view point of common plate theory) fact is originated
from the "wrong" variational procedure, used in GR. From the viewpoint of
deformational approach true variational variables are not Riemannian metric
components $\{g_{\alpha\beta}(x)\},$ but embedding variables $\{y(\xi)\}.$
Varying $\mathfrak{F}$ over $y(\xi)$ we have obtained in \cite{kok3} "right"
plate equilibrium equations for $y(\xi)$ and have proved that they possess more
generality, then Einstein equations.
\item
More detailed analysis shows, that $\mathfrak{F}_b$ is reduced  to an exact
linearized Einstein-Gilbert  action when Poisons coefficient $\sigma_P$ of the
plate is $1/2$. In this case variational derivative of $\mathfrak{F}_{b}$ (over
$g$) transforms into linearized purely geometrical Einstein tensor, whose
divergence vanishes by  Bianchi identities. So, from the viewpoint of the
deformational approach  {\it matter equations of motion follows from the field
equations due to special elastic properties of space-time.}
\item
Curiously, that the Einstein case $\sigma_P=1/2$ is degenerate from the
viewpoint of the deformational approach, since $\mathcal{F}_{0}\,d\mu$ becomes
exact form relatively to embedding variables $x(\xi)$ (but not $g$). In
physical deformational language plate's cylindrical stiffness factors
$\{D_{m}\}_{m=1,\dots N}$ in all $N$ extradimensions vanish under
$\sigma_P=1/2.$
\item
Dimensional manipulations leads to the following relation between Einstein
gravitational constant $\varkappa$ and elastic parameters of $\mathcal{B}$,
supporting old Sacharov's hypothesis \cite{saharov}:
\[
Eh^{N+3}\sim\frac{1}{\varkappa},
\]
where $E$ --- Young modulus of the plate, $N$ --- number of extradimensions,
$h$
--- (averaged) thickness of the plate in extradimensions. Assuming $h\sim
l_{\rm Pl},$ $N\sim1$ we obtain $\ln E(\mbox{\rm Pa})\sim 10^2$
--- huge stiffness of space-time!
\end{itemize}
 Some another interesting topics, involving thermodynamics, origin of hyperbolicity of
 space-time, lagrangian formalism and boundary conditions have been discussed in
 cited papers. Cosmological implication of the theory in the simplest
case $N=1$ has been considered in \cite{kok5}.

\subsection{Example 4: Maxwell electrodynamics as symplectic bimetric
$\Delta^2$-structure.}\label{maxwell}

Let $\mathcal{M}$ --- symplectic  manifold (${\rm dim}\,\mathcal{M}=2n$) with
$\Theta=\omega\in\Lambda^2(\mathcal{M})$ --- symplectic form, which is closed
$(d\omega=0)$ and nondegenerate. As usually we define the mapping $i_z:\
\Lambda^2(\mathcal{M})\to\Lambda(\mathcal{M}), $ where $z\in T\mathcal{M},$ by
the relation:
\begin{equation}\label{i}
i_z\omega(u)\equiv\omega(z,u),
\end{equation}
for all $u\in T\mathcal{M}.$

Let $\mathcal{B}=\mathcal{M}$ and let $F_t(\mathcal{M})$  --- some
diffeomorphism $\mathcal{M}\to\mathcal{M},$ which we consider as a history of
some deformation. It induces corresponding vector field  $A=dF(\widetilde{
d/dt})\in T\mathcal{M}.$ Lets calculate local measure of the deformation. Using
rule of action of Lie derivatives on external forms \cite{warner}:
$\pounds_z=i_z\circ d+d\circ i_z,$ we have:
\[
\dot\omega\equiv\pounds_A\omega=(i_A\circ d+d\circ i_A)\omega
=di_A\omega=d\widehat A=\widehat F,
\]
where closeness of $\omega$ has been used. It is naturally to associate
$\widehat F=d\widehat A$ with {\it Faradey-Maxwell} $2$-form and $\widehat
A=i_A\omega$
--- with {\it electromagnetic potential} $1$-form, whose deformational nature
become clear. Following to the ideology of (unimetric) deformational
structures, $\Delta^2$-structure should be based on the lagrangian:
\[
\mathcal{F}_0^\omega=\mu(F,F)_\omega+\frac{\lambda}{2}({\rm Tr}_\omega(F))^2,
\]
where notations $(\ ,\ )_\omega$ and ${\rm Tr}_\omega$ remind us, that they are
defined relatively to $\omega$ as both $d-$metric and $g$-metric. Easily to
check (for example, using Darboux theorem and going to canonical form of
$\omega:$ $dx^1\wedge dx^2+\dots+dx^{2n-1}\wedge dx^{2n}$), that this
lagrangian is not maxwellian. To get Maxwell electrodynamics we need to
introduce Minkowski metric $\eta$ and, so,  go to  bimetric structure.
Obviously, that ${\rm Tr}_\eta(F)\equiv0,$ and we have:
\[\mathcal{F}_0^\eta=\mu(F,F)_\eta
\]
--- standard maxwellian lagrangian with $\mu=-1/16\pi$ in Gauss units.
Note also,  that $d\mu_\eta=d\mu_\omega\sim\omega^{\wedge n}$.

It is easily to understand the role of
$\mathrm{MOT}_{\mathcal{M}}(\mathcal{M})$ within the considered model. It just
generates  gauge transformations of $\widehat A.$ More exactly, let
$\Phi_t(\mathcal{M})\in \mathrm{MOT}_{\mathcal{M}}(\mathcal{M})$ and let
$v=d\Phi(\widetilde{d/dt}).$ Then by interrelations of
$\mathrm{MOT}_{\mathcal{M}}(\mathcal{M})$ and Hamilton vector fields (see
Sec.\ref{example1} and \cite{arnold}), $v={\rm grad}\, h,$ where $h$ --- some
(local) Hamiltonian  function, generating motion $\Phi_t$ and vector field $v.$
By uniqueness theorem there is isomorphism (up to a constant) between
$\mathrm{MOT}_{\mathcal{M}}(\mathcal{M})$ and set of all Hamilton functions
$\{h\},$  defined by equation $d\Phi(\widetilde{d/dt})={\rm grad}\, h.$ Then we
define {\it gauge mapping}:
\[
\phi_{h}:\ \mathrm{DEF}_{\mathcal{M}}(\mathcal{M})_0\to
\mathrm{DEF}_{\mathcal{M}}(\mathcal{M})_0
\]
by the rule:
\[
d(\phi_h(F))\equiv A_h=dF(\widetilde{d/dt})+{\rm grad}\, h=A+{\rm grad}\, h.
\]
In terms of forms, we'll have
\[
i_{A_h}\omega=i_{A}\omega+i_{{\rm grad} h}\omega=\widehat A+dh
\]
--- gauge transformation of electromagnetic potential.

\vspace{1cm} {\bf Acknowledgements}\\ I am very thankful  to local organizers
of the 5-th Asian-Pacific conference
 for comfortable work conditions and full financial supporting, to
the professor of algebra chair of YSPU A.S.Tikhomirov for valuable
consultations during the preparing of the paper and to E.P.Shtern for technical
supporting.

\appendix

\section{The problem  of $\Theta^{-1}_{\mathcal{B}}$}\label{inverse}

Our  approach to the question of existence of $\Theta^{-1}_{\mathcal{B}}$ will
be based on some well known facts of standard square matrix algebra. Namely, in
case of $d$-metrics, taken as bilinear quadratic forms, we know robust
criteria, which provides  existence of both inverting and scalar density of
weight $-1$:

1) Metric $g$ admits point-wise isomorphism $\Omega^{\otimes 2}\to V^{\otimes
2},$ if and only if ${\rm det}\|g\|\neq0,$ where  $\|g\|$ --- matrix of the
form $g$ in any basis. The element of bivector  space, isomorphic to $g$ will
be $g^{-1}\in V^{\otimes 2},$ which in basis, dual to basis of $\Omega^{\otimes
2} $ has inverse to $\|g\|$ matrix;

2) Let $\|L\|$ --- is matrix of nondegenerate linear transformation in the same
vector space, where form  $g$ is acting.  Then, as well known, matrix of the
form is transformed by the rule:
\begin{equation}\label{transform}
\|g\|=\|L\|^{\rm T}\cdot \|g'\|\cdot \|L\|.
\end{equation}
Taking determinant of the both sides and square root we get: $\left|{\rm
det}\|g'\|\right|^{1/2}=\left|{\rm det}\|g\|\right|^{1/2}/{\rm det}\|L\|$
--- required scalar density of weight $-1.$

Let consider the set $\Omega^{\otimes p}(b)$ of all forms of degree $p$ at some
fixed point $b$ of  $d$-body. The set, after fixing some basis, can be
naturally identified with the space of $p$-cubic real matrices
$\mathrm{M}_{d^{\times p}}$ of dimension $d.$ Let $p=2k,\, k\in\mathbb{N} $ and
let some fixed division of all vector arguments of the $2k-$forms on two set
with  $k$ elements is given. Without loss of generality we can relate the first
$k$ arguments to the first set, and remaining $k$ --- to the second. Let, then,
$\chi$: $(\mathbb{Z}^+_d)^{\times k}\to \{1,2,\dots,d^k\}$
--- some ordering of $(\mathbb{Z}^+_d)^{\times k},$ where
$\mathbb{Z}^+_d$
--- the set of positive integer numbers from $1$ to $d$. This
ordering induces the isomorphism (depending on the ordering) $\chi_\ast$:
$\mathrm{M}_{d^{\times 2k}}\to\mathrm{M}_{d^k\times d^k}$ between spaces of
$p$-cubic matrices  and square matrices of dimension $d^k,$ which maps every
matrix element $A_{\alpha_1\dots\alpha_k\alpha_{k+1}\dots\alpha_{2k}}$ into
matrix element $\chi_\ast(A)_{ab}$ by the following rule\footnote{For example
$a$ and $b$ can be taken as $k-$digits numbers of $d$-adic system of calculus
of respective halfs groups of indexes:
\[a=\alpha_1d^0+\alpha_2d^1+\dots+\alpha_kd^{k-1};\ \  b=
\alpha_{k+1}d^0+\alpha_{k+2}d^1+\dots+\alpha_{2k}d^{k-1}.\]}:
\[
\chi_\ast(A)_{ab}=A_{\chi^{-1}(a)\chi^{-1}(b)}.
\]

The isomorphism lets to pull-back all operations of standard matrix algebra
from $\mathrm{M}_{d^k\times d^k}$ to $\mathrm{M}_{d^{\times 2k}}.$ Namely, let
 the following operations are given on $\mathrm{M}_{d^k\times d^k}$: $\alpha$:
$\mathrm{M}_{d^k\times d^k}\to\mathrm{M}_{d^k\times d^k},$ $\beta$: $
\mathrm{M}_{d^k\times d^k}\to\mathbb{R}$ and $\ast$: $\mathrm{M}_{d^k\times
d^k}\times\mathrm{M}_{d^k\times d^k}\to\mathrm{M}_{d^k\times d^k}.$ Then this
operations by the isomorphism  $\chi_\ast$ induce the operations $\bar \alpha,$
$\bar\beta$ and $\bar\ast$ in $\mathrm{M}_{d^{\times 2k}}$ by the rules: \[
\bar\alpha=\chi_\ast^{-1}\circ\alpha\circ\chi_\ast\ \  (i);\ \
\bar\beta=\beta\circ\chi_\ast\ \ (ii);\ \
\bullet\bar\ast\,\bullet=\chi_\ast^{-1}(\chi_\ast\bullet\ast\chi_\ast\bullet)\
\ (iii).\]

Let $\ast$ --- is standard matrix multiplication  in $\mathrm{M}_{d^k\times
d^k}.$ Then $(iii)$ gives the rule for multiplication of matrices in
$\mathrm{M}_{d^{\times 2k}}$:
\[(A\bar\ast B)_{\alpha_1\dots\alpha_k\alpha_{k+1}\dots\alpha_{2k}}=
\sum\limits_{\beta_1,\dots\beta_k=1}^{d}A_{\alpha_1\dots\alpha_k\beta_1\dots\beta_k}
B_{\beta_1\dots\beta_k\alpha_{k+1}\dots\alpha_{2k}},\] $\forall\ A,B\in
\mathrm{M}_{d^{\times 2k}}.$ Preimage of matrix unit $I_{d^k\times d^k }\equiv
e$ will be matrix $\chi_\ast^{-1}(e)=I_{d^{\times 2k}}\equiv E$ with components
\[E_{\alpha_1\dots\alpha_k\alpha_{k+1}\dots\alpha_{2k}}=
\delta_{\alpha_1\alpha_{k+1}}\delta_{\alpha_2\alpha_{k+2}}\cdots\delta_{\alpha_k\alpha_{2k}}.\]
Let $\alpha$ --- inversion operation in $\mathrm{M}_{d^k\times d^k}.$ Then by
$(i),$ if  inverse matrix exists for image  $\chi_\ast(A),$ it will always
exist for its preimage and $A^{-1}\equiv\chi_\ast^{-1}((\chi_\ast(A))^{-1})$
for every  $A\in\mathrm{M}_{d^{\times 2k}}.$ Let $\beta\equiv{\rm det}.$ Then
$\overline{\rm det}$-operation is well defined in $\mathrm{M}_{d^{\times 2k}}.$
Namely, by $(ii)$ it follows that  $\overline{\rm det} A\equiv{\rm
det}(\chi_\ast(A)).$

Now we clame that the matrix equation
\[A\bar\ast X\equiv A\cdot X=E\] in
$\mathrm{M}_{d^{\times 2k}}$ has solution, if $\overline{\rm det}A\neq0.$ The
solution we call matrix $A^{-1},$ inverse to $A.$ Identifying $E$ in the fixed
coordinate system  with mixed tensor in $(T\mathcal{B})^{\otimes k}\otimes
(T^\ast\mathcal{B})^{\otimes k}$
\[(\mbox{\rm in\ components}\ E^{\alpha_1\ \dots\
\alpha_k}_{\alpha_{k+1}\dots\alpha_{2k}}=\delta^{\alpha_1}_{\alpha_{k+1}}\cdots\delta^{\alpha_k}_{\alpha_{2k}}),\]
$A$ --- with  form\footnote{Here and below we omit for brevity $\|\ \|$ and
identify tensors with matrices, which represent them in some coordinate system.
} $\Theta_\mathcal{B}\equiv (d\iota)^\ast\Theta$ of degree $2k$, we go to the
statement 1 of Sec.\ref{int}.

\section{The problem of  $d\mu$ (internal $d$-structures)}\label{vol}

Let $j$ --- Jacobi matrix of some smooth nondegenerate coordinate
transformation $\xi\to \xi'(\xi)$ on $\mathcal{B}$:
\[j^{\alpha}_{\beta}\equiv\frac{\partial \xi'^\alpha}{\partial \xi^\beta},\]
and  $j^{-1}$ --- its inverting. In the space $\Omega^{\otimes
2k}(\mathcal{B})$ this transformation induces  $2k$-cubic matrix
$J^{-1}\in\mathrm{M}_{d^{\times 2k}},$ such that
\begin{equation}\label{Jacobi}
\Theta'_{\alpha_1\dots\alpha_k\alpha_{k+1}\dots\alpha_{2k}}=
(J^{-1})^{\beta_1\dots\beta_k}_{\alpha_1\dots\alpha_k}
\Theta_{\beta_1\dots\beta_k\beta_{k+1}\dots\beta_{2k}}(J^{-1})^{\beta_{k+1}\dots\beta_{2k}}_{\alpha_{k+1}\dots\alpha_{2k}}.
\end{equation}
Obviously, that   $J^{-1}=(j^{-1})^{\otimes k}.$ The expression (\ref{Jacobi})
has image in $\mathrm{M}_{d^k\times d^k}$:
\[\chi_\ast(\Theta')=\chi_\ast(\Theta)'=\chi_\ast(J^{-1})^{\rm T}\chi_\ast(\Theta)\chi_\ast(J^{-1}),
\] --- the formula similar to the (\ref{transform}). Taking determinant of the
both sides we get:
\begin{equation}\label{det}
{\rm det}\,\chi_{\ast}(\Theta')={\rm \det}\,\chi_\ast(\Theta)[{\rm
det}\,\chi_\ast(J^{-1})]^2=\frac{{\rm det}\,\chi_\ast(\Theta)}{[{\rm
det}\,\chi_\ast(J)]^2},
\end{equation}
where  the relation $\chi_\ast(J^{-1})=(\chi_\ast(J))^{-1}$ has been used,
which, in turn, is direct consequence of the $(i).$ From the (\ref{det}) we
see, that scalar density of weight $-1$ exists  when the expression $[{\rm
det}\,\chi_\ast(J)]^2$ is some degree of ${\rm det}\, j.$ It means, that
degrees of $[{\rm det}\,\chi_\ast(J)]^2$ and ${\rm \det}\, j,$ viewed as
homogeneous polynomial  relatively to derivatives $\partial \xi'/\partial \xi,$
should be connected by the relation:
\begin{equation}\label{fold}
{\rm deg}_{(\partial \xi'/\partial \xi)}[{\rm det}\,\chi_\ast(J)]^2=l\cdot{\rm
deg}_{(\partial \xi'/\partial \xi)}{\rm det} j,
\end{equation}
where $l\in \mathbb{R}.$ Since
\[{\rm deg}_{(\partial \xi'/\partial \xi)}{\rm det} j=d,\
\ {\rm deg}_{\partial \xi'/\partial \xi}[{\rm det}\,\chi_\ast(J)]^2=2\cdot{\rm
deg}\,J\cdot{\rm deg}\,{\rm det }|_{\mathrm{M}_{d^k\times d^k}}=2kd^k,\] we go
to the condition:
\begin{equation}\label{ncond}
l=2kd^{k-1},
\end{equation}
which means, that the expression
\begin{equation}\label{volume}
\left|{\rm det}\,\chi_{\ast}(\Theta)\right|^{1/l}=\left|\overline{\rm
det}\,\Theta\right|^{1/l}=\left|\overline{\rm det}\,\Theta\right|^{1/2kd^{k-1}}
\end{equation}
is the  candidate on the scalar density of weight $-1$ relatively to general
coordinate transformation on $\mathcal{B}.$ As it follows from  (\ref{volume}),
the case of forms of degree 2 is peculiar, since under $k=1$ volume form takes
the standard kind: $|{\rm det}\,\Theta|^{1/2}\,dx^1\wedge\dots\wedge dx^d$ and
dependency on dimension of  $\mathcal{B}$ disappears.

The condition (\ref{fold}) and its consequences (\ref{ncond}) and
(\ref{volume}) are necessary but not sufficient for existence $d\mu,$ since one
should check that the homogeneous polynomial $\left|{\rm
det}\,\chi_\ast(J)\right|^{2/l}$ with right degree $d$ is exactly equal to
${\rm det}\, j.$ Let consider the transformation $\varsigma_{\mu\nu}$:
$j\to\tilde j,$ which  permutates  two lines of  $j$
--- $\nu$-th and $\mu$-th.
The permutation induces the transformation $\bar \varsigma_{\mu\nu}$:
$J\to\tilde J $ in $\mathrm{M}_{d^{\times 2k}},$ which permutates  any matrix
element of $J,$ up indexes of which contains $\mu$ and (or) $\nu$ with the
elements, which have on the same positions  indexes  $\nu$ and (or) $\mu$
respectively. The transformation, in turn, induces transformation
$(\bar\varsigma_{\mu\nu})_\ast$: $\chi_\ast(J)\to\widetilde{\chi_\ast(J)},$
acting by the rule:
$(\bar\varsigma_{\mu\nu})_\ast\chi_\ast(J)\equiv\chi_\ast(\bar\varsigma_{\mu\nu}J).$
It pair-wisely permutates  lines in matrix $\chi_\ast(J)$, whose numbers $a$
has preimages $\chi_\ast^{-1}(a)=\alpha_1\dots\alpha_k,$ containing in their
sequences numbers $\mu$ and (or) $\nu.$ Total number of such permutations in
matrix $\chi_\ast(J)$ is equal:
\begin{equation}\label{perm}
P=\sum\limits_{i=1}^{k}2^{i-1}C^k_i=(3^k-1)/2.
\end{equation}
So, under any permutation of two lines of Jacobi matrix $j,$  (for columns all
statements remains the same), ${\rm det}\,\chi_\ast(J)$ considered as
homogeneous polynomial with respect to  $\partial \xi'/\partial \xi$ is
transformed by the rule ${\rm
det}\,(\bar\varsigma_{\mu\nu})_\ast\chi_\ast(J)=(-1)^P{\rm det}\,\chi_\ast(J).$
It means, that ${\rm det}\,\chi_\ast(J)$ up to a constant factor is $P+2m$-th
($m$
--- any integer) degree of  ${\rm det}\, j,$ which is the unique function
of  $\partial \xi'/\partial \xi$ with required antisymmetry property. By the
kind of isomorphism $\chi_\ast$ (identifying of elements), and by the tensor
product structure $j^{\otimes k}$ of matrix  $J,$ the constant multiplier can
not be dependent on the matrix. The fact, that it is equal unity can be
directly checked  by calculation of determinant of image of identical
coordinate transformation: \[ {\rm det}\,\chi_\ast(E)={\rm det}\, e=+1.\] Now
comparing the expressions \[{\rm det}\,\chi_\ast(J)=({\rm det}\, j)^{P+2m}\]
with (\ref{fold}) and  (\ref{volume}), we get their general consequence:
$P+2m=l/2$ or (using (\ref{perm}) and (\ref{ncond})):
\begin{equation}\label{cond}
3^k-2kd^{k-1}=4m+1,
\end{equation}
which should be considered as equation, relating dimensions of $d$-body and
admissible degree of a $d$-metric within internal $d$-structures. All solutions
of the equation can be pa\-ra\-me\-tri\-zed by the three integer numbers
$(m,k,d).$ For $-5\le m\le5$ there are the following solutions of (\ref{cond}):
\[(0,1,d),\ (0,2,2),\ (1,2,1),\ (-1,2,3),\ (2,2,0),\ (-2,2,4),\]
\[ (-3,2,5),\ (4,4,2),\  (-4,2,6),\ (5,3,1),\ (-5,2,7).\] The first parenthesis says, that forms of
degree 2 can be  $d$-metrics of internal $d$-structures on manifolds with any
dimensions. Easily to check, that for $k=2$ and $k=4$ there are no restrictions
on $d$ too. In case  $k=3$ dimension $d$ can not be even.

\section{Restrictions on  $\Theta$ in $\mathcal{M}$}\label{induce}

To the moment we have considered the form  $\Theta_\mathcal{B},$ as initially
defined on $\mathcal{B}.$ Lets clear what conditions on the form
$\Theta_\mathcal{M}$ and embedding $\iota$ to be satisfied, when
$\Theta_\mathcal{B}=(d\iota)^\ast\Theta$ possesses nondegeneracy property as
induced $d$-metric. We'll use $\chi_\ast$-representation for proving of
statement 3 in Sec.\ref{int}.  Lets turn to the diagram (\ref{diag2}).
\begin{equation}\label{diag2}
\begin{CD}
\Theta_\mathcal{M}(s)@> \ \ \ \ \ \ (d\iota)^\ast(s) \ \ \ \ \ \ \
>>\Theta_\mathcal{B}(b)\\ @V\chi^{(n)}_{\ast}VV @VV\chi^{(d)}_{\ast}V\\
\mathrm{M}_{n^k\times n^k} @>\ \ \   \chi^{(n)(d)}_\ast(d\iota)^\ast\ \ \ \
\
>>\mathrm{M}_{d^k\times d^k}
\end{CD}
\end{equation}

It shows, that after fixing some coordinates system, codifferential
$(d\iota)^\ast$ at every point $s=\iota(b)$ of deformant $\mathcal{S}$ can be
isomorphically represented by the  linear operator\footnote{Now we use
notations $\chi^{(n)}$ and $\chi^{(d)}$ for ordering of different sets
$(\mathbb{Z}^+_n)^{\times k}$ and $(\mathbb{Z}^+_d)^{\times k}$ respectively.}
$\chi^{(n)(d)}_\ast(d\iota)^\ast(s)\equiv\chi_\ast^d\circ
(d\iota)^\ast(s)\circ(\chi_\ast^n)^{-1},$ lying in  ${\rm
Hom}(\mathrm{M}_{n^k\times n^k},\mathrm{M}_{d^k\times d^k}),$ where real linear
spaces of  matrices $\mathrm{M}_{n^k\times n^k}$ and $\mathrm{M}_{d^k\times
d^k}$ represent $\chi_\ast^{(n)}(\Theta_\mathcal{M})$ and
$\chi_\ast^{(d)}(\Theta_\mathcal{B})$ respectively in the fixed basis. In
compact matrix form we have:
\[
\Lambda_b=A^{\rm T}\Lambda_s A,
\]
where $\Lambda_b\equiv\chi_\ast^{(d)}(\Theta_\mathcal{B})(b),$
$\Lambda_s\equiv\chi_\ast^{(n)} (\Theta_\mathcal{M}(s)),$
$A\equiv\chi_\ast^{(nd)}((d\iota)^\ast)$ and $\mathrm{M}_{n^k\times
d^k}\ni\chi_\ast^{(nd)}((d\iota)^\ast)^a_b\equiv((Dx)^{\otimes
k})^{(\chi^{(n)})^{-1}(a)}_{(\chi^{(d)})^{-1}(b)}$ which we can interpret as
element of  ${\rm Hom}(\mathbb{R}^{n^k},\mathbb{R}^{d^k}).$ Here we identify
$\mathbb{R}^{n^k}$ and $\mathbb{R}^{d^k}$ with $(T_s\mathcal{M})^{\otimes k}$
and $(T_b\mathcal{B})^{\otimes k}$ respectively in our fixed coordinate system.
So, we can put the problem at the point in language of real vector spaces. Let
remind some definitions \cite{art}.

Consider $\mathbb{V}_1$ and $\mathbb{V}_2$ --- some real linear vector spaces
and $\mathbb{V}_1^\ast$ and $\mathbb{V}_2^\ast$ --- their dual spaces (of
linear functionals). Let $\Lambda:\
\tm{\mathbb{V}_1}{\mathbb{V}_1}\to\mathbb{R}$ --- bilinear form in
$\mathbb{V}_1.$ The set $L_\Lambda\subset\mathbb{V}_1$ is called {\it left
kernel} of $\Lambda,$ if $\Lambda(l,x)=0$ $\forall\ x\in\mathbb{V}_1$ and
$\forall\ l\in L_\Lambda.$ Similarly, $R_\Lambda\subset\mathbb{V}_2$ is {\it
right kernel} of $\Lambda,$ if $\Theta(x,r)=0$ $\forall\ x\in\mathbb{V}_1$ and
$\forall\ r\in R_\Lambda.$ The form $\Lambda$ is called {\it
nondegenerate\footnote{More generally, the form $\Lambda$ can be {\it
nondegenerate from the left} and {\it from the right}.}}, if
$L_\Lambda=R_\Lambda=0.$ Let $A\in{\rm
Hom}(\mathbb{V}_1^\ast,\mathbb{V}_2^\ast)$ --- some linear mapping of dual
spaces. It has dual conjugated mapping $A^\ast\in{\rm
Hom}(\mathbb{V}_2,\mathbb{V}_1),$ defined by the rule:
\[
(Au)(z)=u(A^\ast z)
\]
for all $u\in\mathbb{V}_1^\ast,$ $z\in\mathbb{V}_2.$ Its kernel
\begin{equation}\label{krn}
\ker A^\ast\equiv\{w\in\mathbb{V}_2\,|\, A^\ast w=0\}.
\end{equation}
 We denote ${\rm
Im}\,A^\ast\equiv V_2^A.$ The mapping $A$ induces mapping $A^2:\
\mathbb{V}_1^{\otimes 2}\to\mathbb{V}_2^{\otimes 2}$ by the rule
\begin{equation}\label{tr}
\Lambda^A(z,w)\equiv A^2\Theta(z,w)=\Theta(A^\ast z,A^\ast w),
\end{equation}
for all $z,w\in\mathbb{V}_2.$ If we fix some basises in $\mathbb{V}_1$ and
$\mathbb{V}_2$, then (\ref{tr}) takes the following matrix form:
\[
\Lambda^A=A^{\rm T}\Lambda A.
\]
When $\Lambda^A$ will be nondegenerate? Let $L_{\Lambda^A}=R_{\Lambda^A}=0,$
then for $z,w\in\mathbb{V}_2$ where $z$ any fixed and  $w$ runs whole
$\mathbb{V}_2$ we have:
\[
\Lambda^A(w,z)=\Lambda^A(z,w)=0\ \Rightarrow\ z=0.
\]
By (\ref{tr}) it means, that for any fixed $z$ and  any $w$ we have:
\[
\Lambda(A^\ast w,A^\ast z)=\Lambda(A^\ast z, A^\ast w)=0\ \Rightarrow\ A^\ast
z=0.
\]
In other words, the narrowing  $\Lambda|_{V_2^A}$ of the form $\Lambda$ must be
nondegenerate. It means, that
\begin{equation}\label{cond1}
(L_\Lambda\cup R_\Lambda)\cap V_2^A=\varnothing.
\end{equation}
Inversely, if $\Lambda|_{V_2^A}$ is nondegenerate, then for any $u$ running
whole $V^A_2$ and for any fixed $v\in V_2^A$ we have
\[\Lambda(u,v)=\Lambda(v,u)=0\ \Rightarrow\ v=0.\]
By (\ref{krn}) we can symbolically express $u$ and $v$ as
\[ u=A^\ast(w+\ker A^\ast),\ v=A^\ast(z+\ker A^\ast),
\]
for some $w,z\in\mathbb{V}_2.$ Then by (\ref{tr}) we have:
\[
\Lambda(A^\ast(w+\ker A),A^\ast(z+\ker A^\ast))=\Lambda^A(w,z)=0\ \Rightarrow\
z\in\ker A^\ast.
\]
Similar conclusion takes place for left kernel of $\Lambda^A.$ So, {\it
necessary (sufficient) condition of $\Lambda^A$ nondegeneracy is nondegeneracy
of $\Lambda|_{V^A_2}$ ($+$ the condition $\ker A^\ast=0.$) }

Now we come back to our initial problem. Identifying:
\[
\mathbb{V}_{1}\equiv (T_s\mathcal{M})^{\otimes k};\ \ \mathbb{V}_2\equiv
(T_b\mathcal{B})^{\otimes k};\ \ A\equiv (d\iota)^\ast;\ \ A^\ast\equiv
d\iota;\ \ V^A_2\equiv (T_s\mathcal{S})^{\otimes k}
\]
and observing, that $\ker d\iota\equiv0,$ since $\iota$ --- embedding, we go to
the expression (\ref{cond1bis}) of Sec.\ref{int}, which is exactly
(\ref{cond1}), where $L_{\Theta(\mathcal{M})}$ and $R_{\Theta(\mathcal{M})}$
are subsets of $V^{\otimes k}(\mathcal{M}),$ such that at every point
$m\in\mathcal{M}$
\[
L_{\Theta(m)}\equiv(\chi_\ast^{(n)})^{-1}(L_{\chi_\ast^{(n)}(\Theta(m))});\ \
R_{\Theta(m)}\equiv(\chi_\ast^{(n)})^{-1}(R_{\chi_\ast^{(n)}(\Theta(m))}).\]

\end{document}